\documentclass[journal]{IEEEtran}
\usepackage{amsthm}
\usepackage{amsfonts}
\usepackage{amssymb}
\usepackage{makecell}
\ifCLASSINFOpdf
  \usepackage[pdftex]{graphicx}
  \usepackage{graphicx}
  \graphicspath{{../pdf/}{../jpeg/}{../png/}}
  \DeclareGraphicsExtensions{.pdf,.jpeg,.png}
\else
  \usepackage[dvips]{graphicx}
  \graphicspath{{../eps/}}
  \DeclareGraphicsExtensions{.eps}
\fi
\ifCLASSOPTIONcompsoc
  \usepackage[caption=false,font=normalsize,labelfont=sf,textfont=sf]{subfig}
\else
\usepackage[caption=false,font=footnotesize]{subfig}
\fi
\usepackage{multirow}
\usepackage{balance}
\usepackage{multicol}
\usepackage{epstopdf}
\usepackage{verbatim} 

\usepackage[english]{babel}
\usepackage{float}
\usepackage{xcolor}
\usepackage[linesnumbered,ruled,vlined]{algorithm2e}

\usepackage[hidelinks]{hyperref}
\usepackage{cite}
\ifCLASSINFOpdf
   \usepackage[pdftex]{graphicx}
   \else
\usepackage[dvips]{graphicx}
\fi
\usepackage[cmex10]{amsmath}
\usepackage[T1]{fontenc}
\newcommand*{\affmark}[1][*]{\textsuperscript{\dag}}

\begin{document}
\title{Bayesian Learning Aided Simultaneous Sparse Estimation of Dual-Wideband THz Channels in Multi-User Hybrid MIMO Systems}

\author{\normalsize{Abhisha~Garg~\IEEEmembership{Graduate Student Member,~IEEE,} Akash~Kumar, Suraj~Srivastava,~\IEEEmembership{Member,~IEEE}, Nimish~Yadav, Aditya~K.~Jagannatham,~\IEEEmembership{Senior Member,~IEEE} and Lajos Hanzo,~\IEEEmembership{Life Fellow,~IEEE}}

\thanks{The work is supported by IEEE SPS scholarship grant for $2023, 2024$ and $2025$. The work of Aditya K. Jagannatham was supported in part by the Qualcomm Innovation Fellowship; in part by the Qualcomm $6$G UR Gift; in part by the Arun Kumar Chair Professorship; and in part by the DST, Govt. of India. Lajos Hanzo would like to acknowledge the financial support of the following Engineering and Physical Sciences Research Council (EPSRC) projects is gratefully acknowledged: Platform for Driving Ultimate Connectivity (TITAN) under Grant EP/Y037243/1 and EP/X04047X/1; Robust and Reliable Quantum Computing (RoaRQ, EP/W032635/1); PerCom (EP/X012301/1); EP/X01228X/1; EP/Y037243/1. The work of S. Srivastava was supported in part by IIT Jodhpur’s Research Grant No. I/RIG/SUS/20240043; in part by Anusandhan National Research Foundation’s PM-ECRG/2024/478/ENS; and in part by Telecom Technology Development Fund (TTDF) under Grant TTDF/6G/368. S. Srivastava, A. K Jagannatham, and L. Hanzo jointly acknowledge the funding support provided to ICON-project by DST and UKRI-EPSRC under India-UK Joint opportunity in Telecommunications Research. 

Abhisha Garg and Aditya K. Jagannatham are with the Department of Electrical Engineering, Indian Institute of Technology Kanpur, Kanpur-$208016$, India (e-mail: abhisha20@iitk.ac.in; adityaj@iitk.ac.in). 

Akash Kumar is with Qualcomm India Pvt. Ltd., Hyderabad, Telangana, $500081$, India (email: akkum@qti.qualcomm.com)

Suraj Srivastava is with the Department of Electrical Engineering, Indian Institute of Technology Jodhpur, Jodhpur, Rajasthan $342030$, India (e-mail: surajsri@iitj.ac.in).

Nimish Yadav is with Samsung Semiconductor India Research, Bengaluru, Karnataka, $560048$, India (email: nimish.y@samsung.com). 

L. Hanzo is with the School of Electronics and Computer Science, University of Southampton, Southampton SO17 1BJ, U.K. (email:lh@ecs.soton.ac.uk)}}
\maketitle
\begin{abstract}
This work conceives the Bayesian Group-Sparse Regression (BGSR) for the estimation of a \textit{spatial} and \textit{frequency wideband}, i.e., a \textit{dual wideband} channel in Multi-User (MU) THz hybrid MIMO scenarios. We develop a practical dual wideband THz channel model that incorporates absorption losses, reflection losses, diffused ray modeling and angles of arrival/departure (AoAs/AoDs) using a Gaussian Mixture Model (GMM). Furthermore, a low-resolution analog-to-digital converter (ADC) is employed at each RF chain, which is crucial for wideband THz massive MIMO systems to reduce power consumption and hardware complexity, given the high sampling rates and large number of antennas involved. The quantized MU THz MIMO model is linearized using the popular Bussgang decomposition followed by BGSR based channel learning framework that results in sparsity across different subcarriers, where each subcarrier has its unique dictionary matrix. Next, the Bayesian Cram{\'e}r Rao Bound (BCRB) is devised for bounding the normalized mean square error (NMSE) performance. Extensive simulations were performed to assess the performance improvements achieved by the proposed BGSR method compared to other sparse estimation techniques. The metrics considered for quantifying the performance improvements include the NMSE and bit error rate (BER).
\end{abstract}
\IEEEpeerreviewmaketitle
\vspace{-5mm}
\section{Introduction}
TeraHertz (THz) communication stands out as a promising avenue to deliver ultra-high data rates and massive short-range connectivity in next-generation wireless communications. The THz band, spaning from $(0.1-10)$ THz, offers spectral windows characterized by bandwidths of $10$ GHz or higher \cite{jornet2011channel}. In comparison to the mmWave band that spans $30-100$ GHz \cite{li2020dynamic}, the major challenges in this band include having fewer spatial degrees-of-freedom (SDoF) \cite{sarieddeen2021overview}, high blockages and the spatial wideband effect. The short wavelength of the THz signal leads to variation in the delay across different antenna elements within the array, resulting in deleterious spatial wideband effects \cite{wang2018spatial}. Furthermore, a wideband THz signal exhibits frequency-selective characteristics stemming from the multipath delay spread, consequently inducing a hostile frequency-wideband effect. A general THz system experiencing simultaneous frequency- and spatial-wideband effects is termed a dual wideband system \cite{garg2024angularly}. The so-called \textit{beam squint} effect arises because of the variation in the effective angle of arrival (AoA)/ angle of departure (AoD) across the subcarriers, which in turn affects the array response vector \cite{wang2018spatial}. This phenomenon leads to a significant difference with respect to its mmWave counterpart and gives rise to a fundamental challenge in this band. However, the presence of precise channel state information (CSI) is essential for successful beamforming in such systems. Furthermore, the conventional channel estimation methods like least-squares (LS) and linear minimum mean squared error (LMMSE) techniques necessitate a substantial pilot overhead because of the dense packing of antennas, leading to reduced spectral efficiency (SE). Therefore, the development of advanced CSI estimation techniques tackling the dual-wideband effect in THz hybrid MIMO systems is paramount, while also accounting for the beam squint effect. A popular technique of recovering such a sparse channel is sparse Bayesian learning (SBL), which exhibits a distinct advantage over other sparse recovery techniques such as LASSO and FOCUSS, as it automatically exploits sparsity, without user intervention \cite{wipf2004sparse}. Additionally, as a probabilistic method, SBL calculates the \textit{a posteriori} distribution of the sparse weight vectors, including its mean and covariance, which enhances the overall accuracy of the estimation process. The following section will offer a critical appraisal of the pertinent previous studies.
\vspace{-2mm}
\subsection{Review of existing works}
The THz band is associated with significant challenges primarily due to the inevitable propagation and molecular absorption (MA) losses. Pioneering research conducted by Jornet and Akyildiz \cite{jornet2011channel} has led to the development of a cutting-edge THz channel model that effectively incorporates both propagation and MA losses. The pioneering work of Lin and Li \cite{lin2015adaptive} proposed a distance- and frequency-dependent hybrid beamforming scheme for an indoor scenario considering multiple users (MU). They modeled the AoA/AoD using a Gaussian Mixture Model (GMM).

The problem of finding sparse solutions from a single measurement vector (SMV) has received extensive attention in prior research \cite{tipping2001sparse}. Despite its NP-hard nature, numerous near-optimal algorithms designed for single-measurement scenarios have been devised, and these algorithms have proven to be valuable in diverse applications \cite{cheng2016matrix}. The algorithms conceived for SMV can broadly be classified into greedy algorithms \cite{huang2018channel}, mixed-norm optimization \cite{du2015mixed}, iterative re-weighted \cite{chethan2016iterative} and Bayesian learning algorithms. As a result of this groundwork, numerous multiple measurement vector (MMV) algorithms have emerged as straightforward extensions of these SMV algorithms. These MMV algorithms leverage the foundational principles established in the SMV context to address more complex scenarios, owing to their efficiency and adaptability. Wipf and Rao \cite{wipf2004sparse} proposed the SBL algorithm for SMV and later on extended to MMV in form of the MMV-SBL (MSBL) algorithm. Within the realm of MMV algorithms, the family of Bayesian learning frameworks has gained significant popularity among researchers due to their demonstrated can be ability of achieving optimal recovery performance. Cotter \textit{et al.} \cite{cotter2005sparse}, in their path breaking work expanded the application of MMV techniques to encompass a range of sparse recovery algorithms, including MMV based matching pursuit (MBMP), MMV orthogonal matching pursuit (MOMP), MMV order-recursive matching pursuit (MORMP), and MFOCUSS with unknown sparsity structure. The key advantage of using MSBL over the other MMV algorithms is that the global minima of MSBL is always the sparsest solution and exhibits very few local minima compared to other MMV based algorithms \cite{srivastava2020msbl}. 

The versatility and relevance of MSBL techniques are not limited to quasi-static channels; they also lend themselves to employment in the domains of online learning and data recovery processes. Joseph \textit{et al.} \cite{joseph2015online}, introduce a pair of online MSBL algorithms designed for recovering temporally correlated sparse data, using a sequential EM approach. They model the correlation using an auto-regressive (AR) process and achieve online estimation with a small delay. Srivastava \textit{et al.} \cite{srivastava2021data}, proposed a MMV sparse Kalman filtering based approach for data-aided channel estimator tailored for doubly selective mmWave MIMO-OFDM systems. They model the temporal variation of the path gain using an AR-$1$ process.

The authors of \cite{srivastava2022hybrid} developed an SBL based TD THz channel estimation framework considering optimal pilot design. Sha and Wang \cite{sha2021channel} in their pioneering work proposed a two-stage channel estimator for frequency wideband THz channels considering RF impairments. Dovelos \textit{et al.} \cite{dovelos2021channel}, modeled the OFDM-based dual-wideband THz channel of a uniform planar array (UPA) and estimated the channel using both OMP and generalized MOMP (GSOMP) based approaches. They consider a single-user (SU) configuration in a single antenna scenario. The recent literature only has a limited number of works addressing the dual wideband effect for THz channel estimation. In this work, Chou \textit{et al.} \cite{chou2023compressed}, proposed a framework for dual wideband channel modeling in the sub-THz regime, utilizing MMV least squares (MMV-LS-CS) techniques for acquiring a time-varying MIMO-OFDM channel. Li and Madhukumar \cite{li2024hybrid} proposed a channel-estimation framework for hybrid (near- and far)-field THz UM-MIMO that integrates dictionary learning with Bayesian sparse recovery. A sparsifying dictionary is learned to capture the joint hybrid-field channel structure, and Bayesian inference is then applied to recover the sparse coefficients, yielding improved estimation performance under constrained pilot resources relative to fixed-basis approaches.

Wang \textit{et al.} \cite{wang2018spatial}, in their groundbreaking treatise suggest a new massive MIMO channel model that considers both the spatial-wideband and frequency-wideband effects. However, they left the problem of dual-wideband reception under low-resolution ADCs as an open issue. Zhang \textit{et al.} \cite{zhang2021analysis}, compare the performance of finite-resolution ADCs and DACs in narrow-band MU THz hybrid MIMO systems. They derived a closed-form expression for the achievable lower bound. Nikbakht and Lozano \cite{nikbakht2021terahertz} proposed an unsupervised learning-based technique for transmit beamforming in THz systems considering low-resolution ADCs. However, none of the existing literature investigating low-resolution ADCs considers the dual-wideband channel, which is predominant in the THz band. The authors in their seminal work \cite{wang2024knowledge} proposed a data-driven de-quantizer to combat hardware imperfections in OFDM systems. They utilized a data-driven approach to mitigate signal distortion caused by quantization and additive white Gaussian noise (AWGN), whereas compressive sensing techniques do not account for system properties. Kim and Choi in their pioneering work \cite{kim2021spatial} proposed a channel estimation scheme considering low-resolution ADCs in the face of spatial-wideband effect only. However, they consider a mmWave massive MIMO system, where each user is equipped with a single antenna. Furthermore, they employ a time-domain system that does not account for frequency selectivity, a common consideration for wideband THz systems. 

Moreover, in contrast to \cite{garg2024angularly}, which primarily focuses on angular-domain processing, this work addresses multiple key challenges inherent in THz communication. Specifically, we incorporate off-grid-based dictionary estimation to enhance the angular resolution, account for low-resolution ADCs which are the dominant sources of power consumption and extension of the framework to construct the MU dual-wideband channel formulation. Additionally, we integrate the GMM based approach for AoA/AoD generation to better capture spatial variations.
\begin{table*}
    \centering
\caption{\small Comparison of the notable contributions of this study with the already existing works\vspace{-0.5\baselineskip}} \label{tab:lit_rev}

\begin{tabular}{|l|c|c|c|c|c|c|c|c|c|c|c|c|c|c|c|}

    \hline

\textbf{Features} & \cite{lin2015adaptive} & \cite{garg2024angularly} &\cite{srivastava2021data}  &  \cite{srivastava2022hybrid} & \cite{dovelos2021channel} & \cite{chou2023compressed} & \cite{li2024hybrid} &\cite{kim2021spatial} &\cite{ding2018bayesian}& \cite{sha2021channel} &\textbf{This Paper} \\

 \hline

sub-THz/ THz Band

& \checkmark & \checkmark &  & \checkmark &  \checkmark & \checkmark & \checkmark &  &  & \checkmark & \checkmark\\

 \hline
 
Reflection losses \& molecular losses

& \checkmark & \checkmark &  &  \checkmark &  \checkmark & \checkmark & \checkmark  &  &  &  & \checkmark\\

 \hline

SC-FDE system

&  & \checkmark &  &  &   &  &  &  &  & \checkmark & \checkmark\\
\hline

Dual-Wideband Effect

&  & \checkmark &  &  & \checkmark  & \checkmark &  & \checkmark &  &  & \checkmark\\
\hline

Multi-User MIMO 

& \checkmark &  &  &  &   &  & \checkmark &  & \checkmark &  & \checkmark\\
\hline

Low Resolution ADCs

&  &  &  &  &   &  & \checkmark & \checkmark & \checkmark &  & \checkmark\\

 \hline

BCRB

&  & \checkmark & \checkmark &  \checkmark & \checkmark  &  & \checkmark &  &  &  & \checkmark\\
 \hline

\textbf{Taylor-based off-grid estimation}

&  &  &  &  &   &  &  &  &  &  & \checkmark\\
 \hline
 
\textbf{MU AoA/AoD with GMM}

&   &  &  &  &   &   &  &  &  &  & \checkmark\\
\hline

\textbf{RRC-PSF based dual-wideband channel}

& &  &  &  &  &  &  & &  &  & \checkmark \\

 \hline
\end{tabular} \vspace{-1 \baselineskip}
\end{table*}
These enhancements collectively provide a more comprehensive and practical framework for robust channel estimation and system design in THz-band communications. To address this research gap, we propose a novel CSI estimation framework that considers a dual wideband channel for MU scenarios using an SC-FDE system. It is important to note that, for low-resolution ADCs, single-carrier systems are generally preferred, since OFDM-based systems cannot preserve orthogonality \cite{mollen2016one}.
Table-\ref{tab:lit_rev} boldly contrasts the salient contributions of the proposed work to the literature at a glance.
\subsection{Contributions}
\begin{enumerate}
    \item We commence by formulating a practical THz channel model that incorporates both frequency- and spatial-wideband effects, while also accounting for their distance-dependent characteristics. Furthermore, we consider the effects of diffused rays associated with each multipath component, a practical challenge that has been overlooked in previous contributions \cite{jornet2011channel}, \cite{lin2015adaptive}, \cite{chou2023compressed}. Additionally, we compare the root raised cosine pulse shaping filter (RRC-PSF) and the rectangular pulse shaping filter (Rect-PSF) in dual-wideband channel formulations. A key challenge is the generation of distinct and realistic spatial signatures for AoA/AoD pairs using the GMM corresponding to each user, which has been overlooked in \cite{lin2015adaptive}, \cite{priebe2011aoa} and addressed in this work.
    \item The approaches in \cite{srivastava2022hybrid} and \cite{gonzalez2018channel} assume a common dictionary across all subcarriers, which is not feasible due to the beam squint effect altering the traditional array manifold. To address this, we leverage the sparsity in the angular domain to formulate a compressive sensing (CS) problem and solve it using the Bayesian Group Sparse Regression (BGSR) approach. Additionally, we construct both the on-grid and Taylor-based off-grid (TBoD) dictionary matrices and demonstrate that the channel exhibits a shared support across different subcarriers, unlike the case in \cite{garg2024angularly}.
    \item We quantify the accuracy of the proposed BGSR based technique by comparing the estimation error variance to the Bayesian Cramér-Rao Bound (BCRB). Therefore, another significant contribution of this work involves deriving the BCRB for MU scenarios. It provides a valuable benchmark for assessing the performance of the CSI frameworks.
    \item Another key contribution of this work involves the combined study of the grave challenges occurring in the THz band, including the hybrid architecture, low-resolution ADCs, and dual-wideband channels, to enhance the end-to-end system performance. To the best of our knowledge, no existing work considers an SC-FDE-based system utilizing low-resolution ADCs in dual-wideband channels. This is because single-carrier systems are inherently more robust to low-resolution ADCs than OFDM systems. Therefore, integrating these elements within a unified SC-FDE framework further exacerbates their interdependencies, which we aim to address in this work.
    \item Our simulation results conclusively illustrate the improved performance of the proposed BGSR algorithm, also verifying that the normalized mean squared error (NMSE) of the proposed algorithm approaches the BCRB, demonstrating its effectiveness. Furthermore, the MA coefficient is evaluated using the HIgh Resolution Transmission (HITRAN) database \cite{rothman2009hitran}, which exhibits reliability throughout the entire THz band.\\
\end{enumerate}
\begin{figure*}
\centering
\subfloat[]{\includegraphics[scale = 0.2]{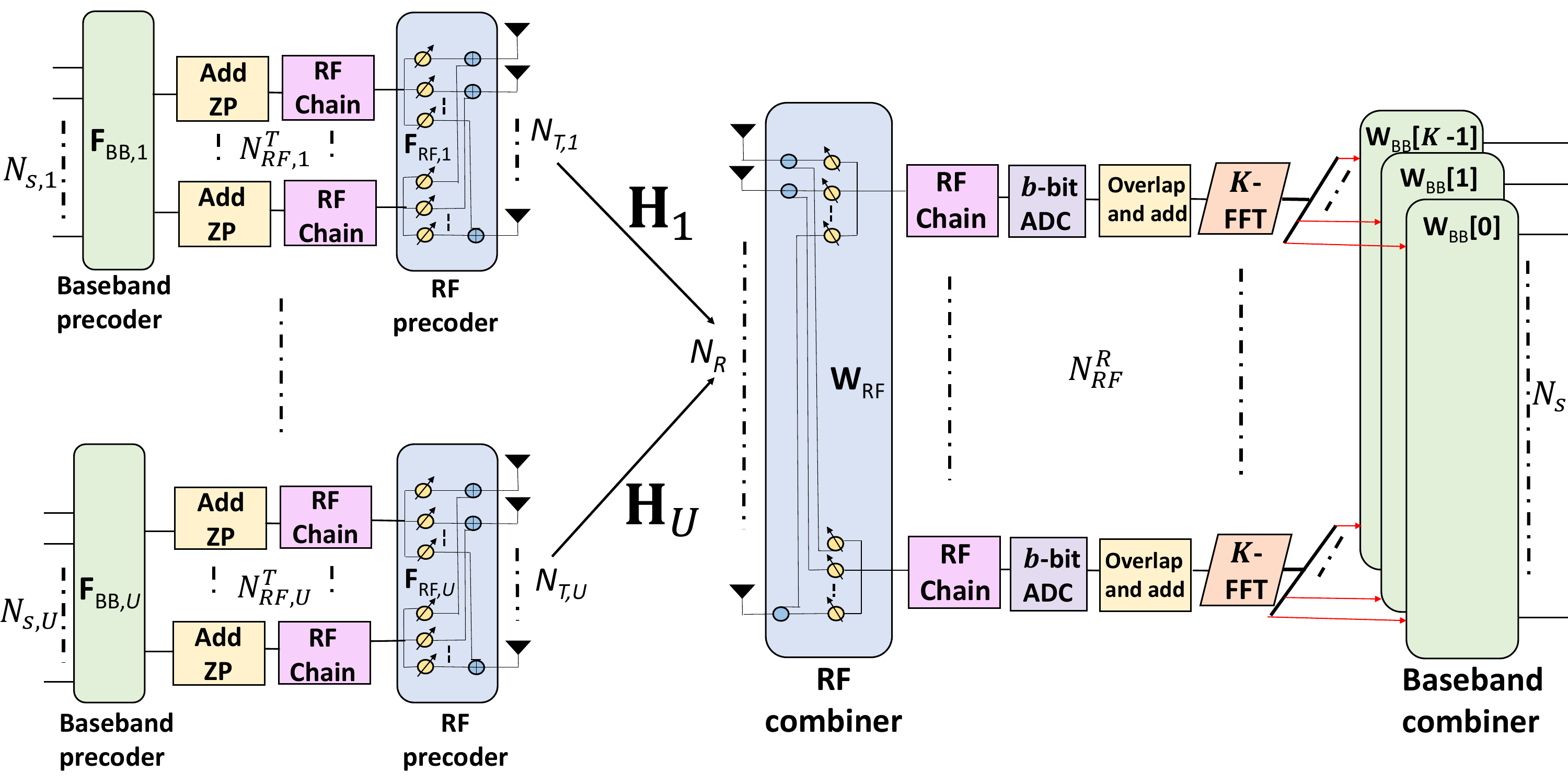}}
\hfil
\hspace{-5pt}\subfloat[]{\includegraphics[scale=0.65]{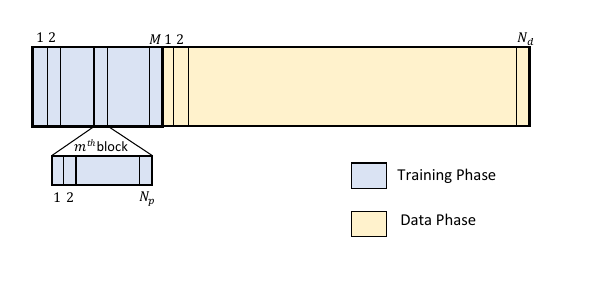}}
\caption{$(a)$ Schematic diagram of SC-FDE based MU THz hybrid MIMO systems with low-resolution ADCs. $(b)$ Frame structure utilized for wideband MU THz hybrid MIMO system using SC-FDE corresponding to each user.}
\label{THz_MIMO} \vspace{-1 \baselineskip}
\end{figure*}
\vspace{-4mm}
\subsection{Notation:} \label{notan} Matrices are represented by uppercase letters $\mathbf{B}$, while lowercase letters $\mathbf{b}$ represent vectors. Various operators are indicated by superscripts such as $(.)^T, (.)^H, (.)^{-1}$ which represent the transpose, Hermitian and inverse, respectively. Let $\left\{\mathbf{H}_{u}(0), \mathbf{H}_{u}(1), \cdots, \mathbf{H}_{u}(N-1)\right\}$ represent a sequence of matrices and $\left\{\mathbf{x}_u(0), \mathbf{x}_u(1), \cdots, \mathbf{x}_u(N-1)\right\}$ represent a sequence of vectors. The circular convolution $\left\{\mathbf{r}_u(n)\right\}_{n=0}^{N-1}$ can be defined as
    \begin{equation}
        \begin{aligned}
        \mathbf{r}_u(n) & = \sum_{l=0}^{N-1} \mathbf{H}_{u}(l)\mathbf{x}_u[(n-l)]_N + \check{\mathbf{v}}_u(n), \notag
    \end{aligned}
    \end{equation}
    where $[.]_N$ represents modulo-N operation. Let $\left\{\mathbf{x}_u(0), \mathbf{x}_u(1), \cdots, \mathbf{x}_u(N-1)\right\}$ represent the input sequence, where $\mathbf{x}(n,k)$ denotes the $k$th element of $\mathbf{x}(n)$. Let $\left\{\mathbf{r}_u(0), \mathbf{r}_u(1), \cdots, \mathbf{r}_u(N-1)\right\}$ constitute the output sequence, where $\mathbf{r}(n,k)$ denotes the $k$th element of $\mathbf{r}(n)$. Then, the N-point FFT of the vector sequence is defined as $\mathbf{r}(p,q) = \sum_{n=0}^{N-1}\mathbf{x}(n,q)e^{-j\frac{2 \pi n p}{N}}$. The uniform distribution is denoted by $\mathcal{U}(.)$ while the Gaussian distribution is denoted with $\mathcal{CN}(\boldsymbol{\mu},\mathbf{\Sigma})$ where $\boldsymbol{\mu}$ represents mean while $\mathbf{\Sigma}$ represents variance.
\section{MU THz Hybrid MIMO System and Channel Estimation Models} \label{MU_model}
Consider an MU THz hybrid MIMO system as shown in Fig. \ref{THz_MIMO}(a), where a BS is fitted with $N_R$ receive antennas and serves $U$ users simultaneously, with each user possessing $N_{T,u}$ transmit antennas. Therefore, the total of all the transmit antennas for all the users equals $\sum_{u=1}^U N_{T,u} = U N_{T,u} = N_T$. The BS has $N_{RF}^R$ RF chains, while each user possess $N_{RF,u}^T$ RF chains. Furthermore, we consider a practical scenario in which the RF chain count follows the constraint $N_{RF}^R \ll N_R$ and $N_{RF,u}^T < N_{T,u}$ in order to support the hybrid MIMO architecture. The aggregate number of data streams, denoted as $N_s$, is subject to the constraint $N_s \leq N_{RF}^R$, where $N_s$ is defined as $N_s = \sum_{u=1}^U N_{s,u} = UN_{s,u}$, and $N_{s,u}$ represents the number of data streams corresponding to each user. Moreover, at the receiver, the hybrid processing module consists of a set of frequency-selective baseband receiver combiners (RC) $\mathbf{W}_{\text{BB}}[k] \in \mathbb{C}^{N_{RF}^R \times N_s}$ cascaded with a frequency-flat RF RC $\mathbf{W}_{\text{RF}} \in \mathbb{C}^{N_R \times N_{RF}^R}$. At the transmitter, the hybrid processing module is comprised of a frequency-flat baseband transmit precoder (TPC) $\mathbf{F}_{\text{BB},u} \in \mathbb{C}^{N_{RF,u}^T \times N_{s,u}}$ for each user coupled with a frequency-flat RF TPC $\mathbf{F}_{\text{RF},u} \in \mathbb{C}^{N_{T,u} \times N_{RF,u}^T}$. Furthermore, the RF TPC and RC are composed of digitally controlled networks of phase-shifters, which consist of only constant-amplitude phase elements following the constraints $\left|\mathbf{W}_{\text{RF}}(\kappa,\ell)\right| = \frac{1}{\sqrt{N_R}}$, $\left|\mathbf{F}_{\text{RF},u}(\kappa,\ell)\right| = \frac{1}{\sqrt{N_{T,u}}}$. 

The frame structure of each user is shown in Fig. \ref{THz_MIMO}(b). Let $\mathbf{H}_{l,u} \in \mathbb{C}^{N_R \times N_{T,u}}$ represent the complex wideband THz MIMO channel corresponding to the $l$-th delay tap of $u$-th user. Furthermore, let $\mathbf{a}_{m,u}^{(p)} \in \mathbb{C}^{N_{RF,u}^T \times 1}$ represent the $p$-th complex pilot vector of the $m$-th block corresponding to the $u$-th user. Prior to transmission through individual RF chains, the $N_p$ pilot vectors undergo zero-padding (ZP), where $L-1$ zeros are appended to each block. This process results in ZP blocks of length $K$, which is further given by $\big\{ {\mathfrak{a}_{m,u}^{(q)}} \big\}_{q=0}^{K-1} = \big\{ {{\mathfrak{a}_{m,u}^{(0)}},{\mathfrak{a}_{m,u}^{(1)}},\cdots,{\mathfrak{a}_{m,u}^{(N_p-1)}},\underbrace{\mathbf{0},\cdots,\mathbf{0}}_{L-1}} \big\}$. To achieve consistent dimensions, $N_p - 1$ zero-matrices of size $N_R \times N_{T,u}$ are appended to the channel taps $\mathbf{H}_{l,u}$, and the resultant taps are given by $\left\{ \mathbf{H}_{\mathit{q},u}\right\}_{\mathit{q}=0}^{\mathit{K}-1} = \Big\{ \mathbf{H}_{0,u},\mathbf{H}_{1,u},\cdots,\mathbf{H}_{\mathit{L}-1,u},\underbrace{\mathbf{0},\cdots,\mathbf{0}}_{\mathit{N}_{p}-1} \Big\}$. Therefore, the received signal vector $\mathbf{r}_m(q) \in \mathbb{C}^{N_{BS} \times 1}$ corresponding to all the $U$ users is given by
\vspace{-2mm}
\begin{align}
    \mathbf{r}_m(q) = \sum_{u=1}^U \mathbf{H}_{q,u} \otimes_K (\mathbf{F}_{\text{RF},m,u} \mathfrak{a}_{m,u}^{(q)}) + \tilde{\mathbf{v}}_{m}(q),
\end{align}
where $\tilde{\mathbf{v}}_{m}(q)$ represents the AWGN, which obeys $\mathcal{CN}(\mathbf{0}_{N_{BS} \times 1}, \sigma^2 \mathbf{I}_{N_{BS}})$, while $\otimes_K$ denotes the circular convolution of length $K$. The quantity $\mathbf{F}_{\text{RF},m,u}$ represents the RF TPC for the $u$-th user corresponding to the $m$-th block. Furthermore, the ZP operation converts the linear convolution to circular convolution and after applying the fast Fourier transform (FFT) at the receiver, the system is decoupled into different frequency bins. At this juncture it is also worth noting that while the overall system model is decomposed into different frequency bins at the receiver, it can be still viewed as an SC system at the transmitter. This is because, the transmitter does not employ an inverse FFT (IFFT) block and the signals $\mathfrak{a}_{m,u}^{(q)}$ are transmitted sequentially with the aid of ZP. Therefore, the received signal $\mathbf{y}_{m,\text{MU}}(q) \in \mathbb{C}^{N_{RF}^R \times 1}$ for all the $U$ users after performing RF combining corresponding to the $m$-th block and passing through low-resolution ADCs $\mathcal{Q}(\cdot)$ can be formulated as
\vspace{-4mm}

\small
\begin{equation}
    \begin{aligned}
    &\mathbf{y}_{m,\text{MU}}(q) = \mathcal{Q}\Big(\mathbf{W}_{\text{RF},m}^H \sum_{u=1}^U\mathbf{H}_{q,u} \otimes (\mathbf{F}_{\text{RF},m,u}\mathfrak{a}_{m,u}^{(q)})+\mathbf{W}_{\text{RF},m}^H\tilde{\mathbf{v}}_m(q)\Big), \\
    & \approx \mathbf{D}\mathbf{W}_{\text{RF},m}^H \sum_{u=1}^U \mathbf{H}_{q,u} \otimes_K (\mathbf{F}_{\text{RF},m,u}\mathfrak{a}_{m,u}^{(q)}) + \mathbf{D}\mathbf{W}_{\text{RF},m}^H\tilde{\mathbf{v}}_m(q) + \check{\mathbf{v}}. \label{multi_freq}
\end{aligned}
\end{equation}
\normalsize
Assuming each RF chain to be quantized independently, the quantization matrix $\mathbf{D} \in \mathbb{C}^{N_{\text{RF}}^R \times N_{\text{RF}}^R}$ may be considered to be diagonal. Moreover, the entries of $\mathbf{D}$ are represented as $\mathbf{D} = \varepsilon\mathbf{I}$, where $\varepsilon = 1-\upsilon$ and $\upsilon$ represents the quantization noise-to-signal power \cite{fan2015uplink}. Let $\mathbb{E}\big\{\mathfrak{a}_{m,u}^{(q)}(\mathfrak{a}_{m,u}^{(q)})^H\big\} = \sigma_b^2 \mathbf{I}_{N_{s,u}}$ where $\sigma_b^2$ represents the pilot power. Additionally, the quantization vector $\check{\mathbf{v}}$ follows the distribution $\mathcal{CN}(\mathbf{0}_{N_{\text{RF}}^R \times 1}, \mathbf{C}_m)$, where $\mathbf{C}_m = \varepsilon(1-\varepsilon)\mathrm{diag}(\mathbf{W}_{\text{RF},m}^H\mathbf{Q}_m\mathbf{W}_{\text{RF},m}+\sigma_n^2\mathbf{W}_{\text{RF},m}^H\mathbf{W}_{\text{RF},m})$. The derivation of this along with that for matrix $\mathbf{Q}_m$ is detailed in Appendix-\ref{noise_cova}. Moreover, the diagonal error covariance matrix results from quantizing each RF chain independently, which is followed from the property of Bussgang decomposition for Gaussian inputs \cite{mezghani2012capacity}. Furthermore, the values of $\upsilon$ are given in Table \ref{bit-resolution} for $b<5$ and can be approximated as $\upsilon = \frac{\pi\sqrt{3}}{2}2^{-2b}$ for $b>5$ \cite{fan2015uplink}. Let $\mathbf{v}_m(q) = \mathbf{D} \mathbf{W}_{\text{RF},m}^H\tilde{\mathbf{v}}_m(q)+\check{\mathbf{v}} \in \mathbb{C}^{N_{\text{RF}}^R \times 1}$ represent the equivalent noise. Therefore, the equivalent noise covariance matrix of $\mathbf{R}_{vv} = \mathbb{E}\{\mathbf{v}_m(q)\mathbf{v}_m^H(q)\} \in \mathbb{C}^{N_{\text{RF}}^R \times N_{\text{RF}}^R}$ can be expressed as $\mathbf{R}_{vv} = \varepsilon^2\sigma_n^2\mathbf{W}_{\text{RF},m}^H\mathbf{W}_{\text{RF},m}+\mathbf{C}_m$.
\begin{table}[]
    \centering
    \vspace{-2mm}
    \caption{$\upsilon$ for different ADC bits $b$ \cite{fan2015uplink}}
    \label{bit-resolution}
    \begin{tabular}{|c|c|c|c|c|c|}
         \hline
        $b$ &  $1$ & $2$ & $3$ & $4$ &$5$\\ \hline
        $\upsilon$ & $0.3634$ & $0.1175$ & $0.03454$ & $0.009497$ & $0.002499$\\ \hline
    \end{tabular} \vspace{-1.5 \baselineskip}
\end{table}
Therefore, the received signal $\mathbf{y}_{m,\text{MU}}[k] \in \mathbb{C}^{N_{RF}^R \times 1}$ corresponding to the $k$-th subcarrier for all the users, obtained via $K$-point FFT ${\left\{\mathbf{y}_{m,\text{MU}}[k]\right\}_{k=0}^{K-1} = \text{FFT}\big(\left\{\mathbf{y}_{m,\text{MU}}(q)\right\}_{q=0}^{K-1}\big)}$ is given by
\begin{table*}[]
    \centering
    \vspace{-2mm}
    \caption{Notation and description of channel parameters considered}
    \label{notatn}
    \resizebox{0.8\textwidth}{!}{
    \begin{tabular}{|c|l|c|r|} 
        \hline
        \textbf{Parameter} &  \textbf{Description} & \textbf{Parameter} & \textbf{Description}\\ \hline
        $\lambda$ &  wavelength & $\varpi$ & phase of complex path gain\\ \hline
        $f_k$ &  subcarrier frequency & $K_{\text{abs}}$ & molecular-absorption loss\\ \hline
        $\triangle \vartheta$ & spatial phase offset & $\Upsilon$ & first-order reflection coefficient\\ \hline
        $d_c$ & inter-antenna spacing & $\varrho_k$ &  relative subcarrier frequency \\ \hline
        $N_{ray}$ & $\#$ of diffused rays & $\mu_0/ \epsilon_0$ & free-space permeability/ permittivity \\ \hline
        $\phi_{(.)}$ & AoA & $\varkappa$ & absorption coefficient of reflecting medium \\ \hline
        $\theta_{(.)}$ & AoD & $L_{\text{free}}$ & free-space loss \\ \hline
        $L_{\text{abs}}$ & absorption loss & $Z_0$ & intrinsic impedance of free space \\ \hline
        $\alpha_{(.)}$ & complex path gain & $\nu_i/ \nu_r$ & angle of incidence/ refraction of medium\\ \hline
        $T_s$ & sampling time & $Z$ & characteristic impedance \\ \hline
        $\tau_{(.)}$ & delay & $\eta$ & index of refraction \\ \hline
        $p_{(.)}$ & pulse-shaping filter & $\sigma_r$ & standard roughness deviation of reflecting medium \\ \hline
    \end{tabular}  }
    \\
        \small{Note that, the notation $(\cdot)$ represents a generalized index placeholder corresponding to specific parameter, depending on the context.} \vspace{-1.2 \baselineskip}
\end{table*}
\begin{align}
    \mathbf{y}_{m,\text{MU}}[k] \approx \mathbf{D}\mathbf{W}_{\text{RF},m}^H \mathbf{H}_{\text{MU}}[k]\mathbf{F}_{\text{RF},m,u}\mathfrak{a}_{m,u}[k] + \mathbf{v}_m[k], \label{buss_out}
\end{align}
where $\mathbf{H}_{\text{MU}}[k] = \left[\mathbf{H}_1[k] \; \mathbf{H}_2[k] \; \cdots \; \mathbf{H}_U[k]\right] \in \mathbb{C}^{N_R \times N_T}$ represents the concatenated channel for all the users while $\mathbf{H}_u[k] \in \mathbb{C}^{N_R \times N_{T,u}}$ represents the $K$-point FFT of $\{\mathbf{H}_{l,u}\}_{l=0}^{L-1}$. The quantity $\mathfrak{a}_{m,u}[k] \in \mathbb{C}^{N_{RF,u}^T \times 1}$ is obtained by taking the $K$-point FFT of the input signal $\{\mathfrak{a}_{m,u}^{(q)}\}$ whereas $\mathbf{v}_m[k] \in \mathbb{C}^{N_{RF}^R \times 1}$ results from applying the same transformation to the effective noise signal $\mathbf{v}_m(q)$. Let $\mathbf{s}_{m,\mathrm{MU}}[k]=\mathbf{F}_{\mathrm{RF},m,u}\mathfrak{a}_{m,u}[k] \in \mathbb{C}^{N_{T,u} \times 1}$ for notational simplicity. Upon employing the $\mathrm{vec}(.)$ operator property in Eq. \eqref{buss_out}, we obtain
\begin{align}
    \mathbf{y}_{m,\mathrm{MU}}[k] = \underbrace{(\mathbf{s}^T_{m,\mathrm{MU}}[k] \otimes \mathbf{D} \mathbf{W}_{\text{RF},m}^H)}_{\boldsymbol{\Lambda}_{m,
    \mathrm{MU}}[k]} \underbrace{\mathrm{vec}(\mathbf{H}_{\text{MU}}[k])}_{\mathbf{h}_{\mathrm{MU}[k]}} + \mathbf{v}_m[k]. \notag
\end{align}
Note that, although the Bussgang approximation is sub-optimal under low-resolution ADCs, it remains a practical choice in THz systems \cite{sarieddeen2021overview}. Due to the extremely high bandwidth and large antenna arrays, exact nonlinear quantization models become computationally expensive and scale poorly with system dimensions. However, Bussgang linearization provides a tractable \textit{trade-off} sacrificing some modeling accuracy at higher resolutions, while enabling large-scale multi-subcarrier, multi-antenna channel estimation in THz systems, where beam-squint and high dimensionality already render exact quantization approaches intractable. The next section will discuss the dual-wideband effected channel model. 
\section{MU THz Channel Modeling with dual-wideband effects} \label{MU_channel}
This section delves into the characteristics of the dual-wideband THz MIMO channel for a generalized pulse shaping filter (PSF) and further describes the channel model for a THz hybrid MIMO system. Table-\ref{notatn} provides the description of the channel parameters considered. The generalized array steering vector $\mathbf{a}(\phi,f_c)$ can be formulated as
\begin{align}
    \mathbf{a}(\phi,f_c) = \frac{1}{\sqrt{N}}\Big[1, e^{-j\frac{2 \pi}{\lambda}d_c \cos{\phi}, \cdots, e^{-j(N-1) \frac{2 \pi}{\lambda} d_c \cos{\phi}} }\Big]^T, \label{normal_array_response}
\end{align}
where $\lambda = \frac{c}{f}$ while $d_c = \frac{c}{2f_c}$. It is worth noting that the frequently employed approximation of $d_c \approx \frac{\lambda}{2}$ used for antenna arrays holds true only for narrowband channels, where the $B$ is much smaller than the center frequency $f_c$, i.e., bandwidth $B \ll f_c$. Naturally, it breaks down for wideband signals occupying a significant frequency spectrum \cite{liu2018hybrid}. Let $\mathit{f}_{\mathit{k}}$ represent the frequency of the $\mathit{k}$th subcarrier, which is defined
\begin{align}
f_k \triangleq f_c + \Big(k - \frac{K + 1}{2}\Big)\frac{B}{K}. \label{sub_freq}
\end{align}
Furthermore, we define the spatial phase offset $\triangle\vartheta_{n,k}$ for the $n$-th antenna at the $k$-th subcarrier as
\begin{align}
    \triangle\vartheta_{n,k} = \frac{2 \pi}{\lambda_k} (n-1) d_c \cos{\phi} = \pi (n-1) \varrho_k \cos{\phi},
\end{align}
where $\varrho_{k} = \frac{f_k}{f_c}$. Consequently, the effective spatial AoA $\widehat{\phi}(f_k)$ at the $k$-th subcarrier is defined as
\begin{align}
    \widehat{\phi}(f_k) = \arccos{(\varrho_k \cos{\phi})}. \label{spatial_effect}
\end{align}
Thus, the modified array steering vector associated with frequency-dependent angles from Equations \eqref{normal_array_response} and \eqref{spatial_effect}, can be further expressed as
\begin{align}
    \tilde{\mathbf{a}}(\phi, f_k) = \frac{1}{\sqrt{N}} \left[1, e^{-j\pi \varrho_k \cos({\phi})}, \cdots, e^{-j(N-1) \pi \varrho_k \cos{(\phi})} \right]^T.
\end{align}
Moreover, the THz channel can be characterized as a combination of line-of-sight (LoS) and non-line-of-sight (NLoS) components. The LoS component captures the direct propagation path between the BS and UE, while the NLoS component encompasses the multiple reflections caused by various scatters within the environment. Therefore, the dual-wideband THz MIMO channel can be mathematically expressed as the merger of the LoS and NLoS components, which can be described for the subcarrier $\mathit{k}$ as
\vspace{-2mm}
\begin{align}
\mathbf{H}_u[k] = \mathbf{H}_u^{\text{LoS}}[k] + \mathbf{H}_u^{\text{NLoS}}[k]. \label{Hsub_Los_NLos}
\end{align}
The LoS and NLoS components for the dual-wideband channel are given by
\vspace{-2mm}
\begin{align}
\mathbf{H}^{\textrm{LoS}}_u[{\mathit{k}}] = {\sqrt{{N_{T,u}N_R}}}\alpha(f_k,d)\beta_{\tau}& \mathfrak{B}_{T,u} \mathfrak{B}_R \tilde{\mathbf{a}}_R(\phi,f_k)\tilde{\mathbf {a}}_{T,u}^{H}(\theta,f_k), \label{HLos_sub}
\end{align}
\vspace{-8mm}
\begin{align}
\mathbf{H}^{\textrm{NLoS}}_u[k] = & {\sqrt{\frac{N_{T,u}N_R}{N_{\textrm{NLoS}}N_{ray}}}}\sum_{z=1}^{N_{\textrm{NLoS}}} \sum_{\jmath=1}^{N_{ray}} \alpha_{z,\jmath}(f_k, d_{z,\jmath}) \beta_{\tau_{z,\jmath}} \notag \\ &  \mathfrak{B}_{T,u} \mathfrak{B}_R \tilde{\mathbf{a}}_R(\phi_{z,\jmath},f_k)\tilde{\mathbf{a}}_{T,u}^H(\theta_{z,\jmath},f_k), \label{NLoS_channel}
\end{align}
\vspace{-7mm}
\begin{align}
\beta_{\tau_{z,\jmath}} = \sum_{l=0}^{K-1}p(lT_s-\tau_{z,\jmath})e^{-j\frac{2\pi k l}{K}},  \forall \, k,l, \label{pulse_shaping}
\end{align}
where the parameters of $\mathbf{H}^{(.)}_u[k]$ are defined in Table-\ref{notatn}. Moreover, THz wireless channels are susceptible to substantial attenuation due to various factors. Among these are absorption losses arising from molecular interactions and noise introduced by water vapor molecules \cite{jornet2011channel}. These considerations are integrated into the modeling of the complex-path gains $\alpha(f_k, d)$ and $\alpha_{z,\jmath}(f_k, d_{z,\jmath})$, which depend on both the subcarrier frequency $f_k$ and transmission distance $d_{(.)}$. Moreover, the intricate path gain $\alpha_{(.)}(f_k,d_{(.)})$, can be described by its magnitude and phase components as $\alpha_{(.)}(f_k,d_{(.)}) = |\alpha_{(.)}(f_k,d_{(.)})|e^{j \varpi}$. Thus, the modeling of the complex path gain for the LoS component as discussed in \cite{jornet2011channel} is given by
\begin{align}
    |\alpha(f_k, d)|^2 = L_{\text{free}}(f_k, d) L_{\text{abs}}(f_k, d), \label{LoS_component}
\end{align}
where $L_{\text{free}}(f_k, d)$ and $L_{\text{abs}}(f_k, d)$ \cite{chou2023compressed} are respectively given as
\begin{align}
    L_{\text{free}}(f_k, d) = \left(\frac{c}{4 \pi f_k d}\right)^2, L_{\text{abs}} = e^{-k_{\text{abs}}(f_k)d},
\end{align}
while $k_{\text{abs}}(f_k) = \mathop{\sum}\limits_{p,q}k_{\text{abs}}^{p,q}(f_k)$. The quantity $k_{\text{abs}}^{p,q}(f_k)$ represents the attenuation of radiation at the subcarrier frequency $f_k$ due to the absorption by the $p$-th isotopologue of the $q$-th gas and can be efficiently calculated from the HITRAN database \cite{jornet2011channel}. In a similar vein, the complex path gain associated with the $z$-th NLoS component corresponding to the $\jmath$-th diffuse ray can be mathematically modelled as
\vspace{-2mm}
\begin{align}
    |\alpha_{z,\jmath}(f_k, d_{z,\jmath})|^2 = \Upsilon^2_{z,\jmath}(f_k) L_{\text{free}}(f_k, d_{z,\jmath}) L_{\text{abs}}(f_k, d_{z,\jmath}) \label{NLoS_component}.
\end{align}
Here, $\Upsilon^2_{z,\jmath}$ signifies the first-order reflection coefficient pertaining to the $\jmath$-th diffuse ray within the $z$-th NLoS cluster. The quantity $\Upsilon^2_{z,\jmath}$ is the multiplication of the Fresnel reflection coefficient and Rayleigh roughness factor \cite{piesiewicz2007scattering}, as
\begin{align}
    \Upsilon^2_{z,\jmath} = \frac{Z(f_k) \cos({\nu}_{i_{z,j}}) - Z_0 (\cos{\nu}_{r_{z,\jmath}})}{Z(f_k) \cos({\nu}_{i_{z,j}}) + Z_0 (\cos{\nu}_{r_{z,j}})} e^{-\frac{1}{2}\left(\frac{4 \pi f_k \sigma_{r} \cos\left({\nu}_{i_{z,\jmath}}\right)}{c}\right)^2}, \notag
\end{align}
where ${\nu}_{r_{z,\jmath}} = \arcsin\left(\sin\left({\nu}_{i_{z,\jmath}}\right)\frac{Z(f_k)}{Z_0}\right)$. The characteristic impedance $Z(f_k)$ of the reflecting medium given by $Z(f_k) = \sqrt{\frac{\mu_0}{\epsilon_0 \big(\eta^2 - (\frac{\varkappa c}{4 \pi f_k}) - j\frac{2 \eta \varkappa c}{4 \pi f_k} \big)}}$ where the specific parameter values are detailed in Table-\ref{materials_para}. It is worth noting that the highly directional nature of propagation in the THz
 band results in a sparsely populated multipath channel in the AoA/AoD domain. In such scenarios, the recently developed sparse signal processing paradigm can yield excellent results for channel estimation, which leads to improved bandwidth efficiency. The next section introduces the sparse THz recovery model.
\section{Sparse Channel Estimation for MU THz hybrid MIMO systems} \label{MU_CSI_estimate}
Let $G_{T,u}$ and $G_R$ represent the number of transmit and receive angular bins, satisfying the relationship $(G_{T,u},G_R) \geq \text{max}(N_{T,u}, N_R)$. Furthermore, let $\Theta_{T,u}$ and $\Phi_R$ represent the transmit and receive angular grids, which are formed by the directional cosines in the range $[-1,1]$ and are given by
\vspace{-2mm}
\begin{align}
    \Phi_R &= \big\{\phi_r: \text{cos}(\phi_r) = \frac{2}{G_R}(r-1)-1, 1 \leq r \leq G_R \big\}, \notag \\
    \Theta_{T,u} &= \big\{\theta_{t,u}: \text{cos}(\theta_{t,u}) = \frac{2}{G_{T,u}}(t-1)-1, 1 \leq t \leq G_{T,u} \big\}.
\end{align}
The extended virtual channel model \cite{rodriguez2018frequency} for to the $u$-th user is given by
\vspace{-4mm}
\begin{align}
    \mathbf{H}_u[k] = \mathbf{A}_{R}[k] \mathbf{H}_{b,u}[k] \mathbf{A}_{T,u}^H[k], \label{virtchannel}
\end{align}
where $\mathbf{A}_{R}(\Phi_R, f_k) \in \mathbb{C}^{N_R \times G_R}$ represents the receive array manifold which is given by
\vspace{-2mm}
\begin{align}
    \mathbf{A}_R(\Phi_R, f_k) = \left[\tilde{\mathbf{a}}_R(\phi_1,f_k),\tilde{\mathbf{a}}_R(\phi_2,f_k),\cdots,\tilde{\mathbf{a}}_R(\phi_{G_R},f_k)\right]. \label{receive_array}
\end{align}
The quantity $\mathbf{A}_{T,u}(\Theta_{T,u}, f_k) \in \mathbb{C}^{N_{T,u} \times G_{T,u}}$ denotes the transmit array manifold vector for the $u$-th user, which is given
\vspace{-2mm}
\begin{equation}
    \begin{aligned}
    \mathbf{A}_{T,u}(\Theta_{T,u}, f_k) = \big[\tilde{\mathbf{a}}_{T,u}(\theta_{1,u}, f_k), & \tilde{\mathbf{a}}_{T,u}(\theta_{2,u}, f_k), \cdots, \\ & \tilde{\mathbf{a}}_{T,u}(\theta_{G_{T,u}}, f_k)\big]. \label{trans_grid}
\end{aligned}
\end{equation}
Furthermore, $\mathbf{H}_{b,u}[k] \in \mathbb{C}^{G_R \times G_{T,u}}$ represents the beamspace channel frequency response (CFR) matrix corresponding to $\mathbf{H}_u[k]$. Leveraging the $\text{vec}(.)$ operator and the $\otimes$ relationship, the THz MIMO channel corresponding to the $u$-th user can be recast as
\vspace{-2mm}
\begin{align}
    \text{vec}(\mathbf{H}_u[k]) = \underbrace{\left(\mathbf{A}_{T,u}^{*}(\Theta_{T,u},f_k) \otimes \mathbf{A}_R(\Phi_R, f_k)\right)}_{\tilde{\mathbf{\Psi}}_u[k]} \text{vec}(\mathbf{H}_{b,u}[k]), \notag
\end{align}
where $\tilde{\mathbf{\Psi}}_u[k] \in \mathbb{C}^{N_R N_{T,u} \times G_R G_{T,u}}$ represents the \textit{sparsifying-dictionary} of the $u$-th user. Furthermore, with $\text{vec}\left(\mathbf{H}_{b,u}[k]\right)$ denoting $\mathbf{h}_{b,u}[k]$, the concatenated vectorized CFR across all the $U$ users for the $k$-th subcarrier is given by $\mathbf{h}_{\text{MU}}[k] = \text{blkdiag}\underbrace{\big(\Tilde{\mathbf{\Psi}}_1[k] \: \Tilde{\mathbf{\Psi}}_2[k] \: \cdots\: \Tilde{\mathbf{\Psi}}_U[k]\big)}_{\mathbf{\Psi}_{\mathrm{MU}}[k]}\underbrace{\big[\mathbf{h}_{b,1}^T[k] \mathbf{h}_{b,2}^T[k]  \cdots \mathbf{h}_{b,U}^T[k]\big]^T}_{\mathbf{h}_{b,\text{MU}}[k]},$
where $\mathbf{\Psi}_{\text{MU}}[k] \in \mathbb{C}^{U N_R N_{T,u} \times G_R \sum_{u=1}^U G_{T,u}}$ represents the concatenated sparsifying dictionary and $\mathbf{h}_{b,\text{MU}}[k] \in \mathbb{C}^{G_R \sum_{u=1}^U G_{T,u} \times 1}$ is the concatenated beamspace output for all the $U$ users. Consider a simplistic scenario associated with $\sum_{u=1}^U G_{T,u} = UG_{T,u}$. The beamspace received output vector for all the $U$ users corresponding to the $m$-th block is given by
\vspace{-3mm}
\begin{align}
    \mathbf{y}_{m,\text{MU}}[k] = \underbrace{\mathbf{\Lambda}_{m,\text{MU}}[k]\mathbf{\Psi}_{\text{MU}}[k]}_{\mathbf{\Omega}_{m,\text{MU}}[k]}\mathbf{h}_{b,\text{MU}}[k] + \mathbf{v}_m[k]. \label{m_final}
\end{align}
Furthermore, to develop a compatible pilot model \cite{gonzalez2018channel} for the $k$-th subcarrier, we concatenate the outputs $\mathbf{y}_{m,\text{MU}}[k]$ for all the $M$ blocks into a single vector, represented as
\vspace{-2mm}
\begin{align}
    \underbrace{\begin{bmatrix}
        \mathbf{y}_{1,\text{MU}}[k] \\ \mathbf{y}_{2,\text{MU}}[k] \\ \vdots \\ \mathbf{y}_{M,\text{MU}}[k]
    \end{bmatrix}}_{\mathbf{y}_{\text{MU}}[k]} = \underbrace{\begin{bmatrix}
        \mathbf{\Omega}_{1,\text{MU}}[k] \\ \mathbf{\Omega}_{2,\text{MU}}[k] \\ \vdots \\ \mathbf{\Omega}_{M,\text{MU}}[k]
    \end{bmatrix}}_{\mathbf{\Omega}_{\text{MU}}[k]} \mathbf{h}_{b,\text{MU}}[k] + \underbrace{\begin{bmatrix}
        \mathbf{v}_1[k] \\ \mathbf{v}_2[k] \\ \vdots \\ \mathbf{v}_M[k]
    \end{bmatrix}}_{\mathbf{v}_{\text{MU}[k]}}, \label{final_system}
\end{align}
where $\mathbf{y}_{\text{MU}}[k] \in \mathbb{C}^{MN_{RF}^R \times 1}$ denotes the stacked pilot output, $\mathbf{\Omega}_{\text{MU}}[k] \in \mathbb{C}^{M N_{RF}^R \times UG_RG_{T,u}}$ is the equivalent sensing matrix and $\mathbf{v}_{\text{MU}}[k] \in \mathbb{C}^{M N_{RF}^R \times 1}$ represents the equivalent noise across all the $M$ blocks for the $k$-th subcarrier. We assume that the noise samples are uncorrelated across all the subcarriers and satisfy $\mathbb{E}\left\{\mathbf{v}_{\text{MU}}[k](\mathbf{v}_{\text{MU}}[k])^H\right\} = \text{blkdiag}\left(\mathbf{R}_{vv,1}, \mathbf{R}_{vv,2}, \cdots, \mathbf{R}_{vv,M} \right) = \mathbf{C}_w$. The next section introduces a BGSR based algorithm for the sparse channel estimation model of \eqref{final_system}.
\section{Bayesian Group-Sparse Regression algorithm for sparse channel estimation in MU THz MIMO systems} \label{G-MSB}
As discussed in Section-I, practical scenarios in the THz domain necessitate careful consideration and modeling of the beam squint effect, resulting from the variation in the effective AoA/AoDs across subcarriers. This variation further impacts the array response dictionaries \cite{dovelos2021channel}. Consequently, conventional MSBL algorithms, which assume a common dictionary matrix across all the subcarriers \cite{srivastava2021data}, do not lead to optimal performance in such scenarios. Therefore, we propose a novel BGSR based approach that jointly processes the hyperparameters across all the subcarriers and provides enhanced performance over the conventional sparse sensing techniques. Furthermore, in order to construct a group based MU sparse channel estimation model, we concatenate the output across all the subcarriers from Equation \eqref{final_system} to obtain $\mathbf{Y}_{\text{MU}} = \left[\mathbf{y}_{\text{MU}}[0] \: \mathbf{y}_{\text{MU}}[1]\: \cdots \: \mathbf{y}_{\text{MU}}[K-1] \right] \in \mathbb{C}^{MN_{RF}^R \times K}$. Furthermore, the equivalent sensing matrix across all subcarriers can be given by $\mathbf{\Xi}_{\text{MU}}(:,:,k) = \mathbf{\Omega}_{\text{MU}}[k] \in \mathbb{C}^{MN_{RF}^R \times U G_R G_{T,u} \times K} \: \forall \: 0 \leq k \leq K-1$, and the corresponding noise matrix can be concatenated as $\mathbf{V}_{\text{MU}} = \left[\mathbf{v}_{\text{MU}}[0] \: \mathbf{v}_{\text{MU}}[1] \: \cdots \: \mathbf{v}_{\text{MU}}[K-1]\right] \in \mathbb{C}^{MN_{RF}^R \times K}$. Let $\mathbf{H}_{b,\text{MU}} = \left[\mathbf{h}_{b,\text{MU}}[0] \: \mathbf{h}_{b,\text{MU}}[1] \: \cdots \mathbf{h}_{b,\text{MU}}[K-1] \right] \in \mathbb{C}^{UG_RG_{T,u} \times K}$ represent the aggregate channel matrix across all the subcarriers, while $\mathbf{H}_{b,u} = \left[\mathbf{H}_{b,u}[0] \: \mathbf{H}_{b,u}[1] \: \cdots \mathbf{H}_{b,u}[K-1] \right] \in \mathbb{C}^{G_R \times K G_{T,u}}$ represent the combined channel matrix across all the subcarriers of the $u$-th user. Moreover, we assign a parameterized Gaussian prior $f(\mathbf{H}_{b,\text{MU}};\mathbf{\Gamma}_{\text{MU}})$ to the beamspace channel matrix $\mathbf{H}_{b,\text{MU}}$, which is given by

\small
\begin{equation}
    \begin{aligned}
    f(\mathbf{H}_{b,\text{MU}}; \mathbf{\Gamma}_{\text{MU}}) = \prod_{k=0}^{K-1} \prod_{i=1}^{UG_RG_{T,u}} (\pi \gamma_{k,i})^{-1} \text{exp}\left(-\frac{|\mathbf{H}_{b,\text{MU}}(i,k)|^2}{\gamma_{k,i}}\right), \label{pri_or}
\end{aligned}
\end{equation}
\normalsize
where $\gamma_{k,i}$ represents the $i$-th hyperparameter for the $k$-th subcarrier.   \begin{align}
    \widehat{\mathbf{H}}_{b,\text{MU}}(:,k) & = \left(\mathbf{\Xi}^H_{\text{MU}}(:,:,k)\mathbf{C}_w^{-1}\mathbf{\Xi}_{\text{MU}}(:,:,k) + \mathbf{\Gamma}_{k,\text{MU}}^{-1}\right)^{-1} \notag \\ &\mathbf{\Xi}_{\text{MU}}^H(:,:,k)\mathbf{C}_w^{-1}\mathbf{Y}_{\text{MU}}(:,k) \: \forall \: 0 \leq k \leq K-1, \label{errorcov_MU}
\end{align}
\begin{align}
    \prod_{k=0}^{K-1} & \text{log}[f(\mathbf{Y}_{\text{MU}}(:,k);\mathbf{\Gamma}_{k,\text{MU}})] = \sum_{k=0}^{K-1} \big( -MN_{RF}^R \text{log}(\pi) - \notag \\ & \text{log}[\text{det}(\mathbf{C}_{w_{y,k}})]  - \mathbf{Y}_{\text{MU}}^H(:,k)\mathbf{C}_{w_{y,k}}^{-1}\mathbf{Y}_{\text{MU}}(:,k) \big), \label{MU_log}
\end{align}
The matrix $\mathbf{\Gamma}_{\text{MU}}$ can be defined as $\mathbf{\Gamma}_{\text{MU}} = \sum_{k=0}^{K-1} \mathbf{\Gamma}_{k,\text{MU}} \in \mathbb{R}^{UG_RG_{T,u} \times UG_RG_{T,u}}$, where $\mathbf{\Gamma}_{k,\text{MU}} = \text{diag}\left(\gamma_{k,1},\gamma_{k,2},\cdots,\gamma_{k,UG_RG_{T,u}} \right)$. In Bayesian techniques, hyperparameters are used as variance parameters that play a crucial role in \textit{promoting sparsity,} since coefficients associated with small $\gamma_i$ are automatically driven toward zero, while those with large $\gamma_i$ remain active \cite{tipping2001sparse}. This mechanism eliminates the need to explicitly impose sparsity levels or tune penalty factors, as done in traditional approaches such as OMP or LASSO \cite{wipf2007empirical}. Furthermore, by iteratively updating the hyperparameters, Bayesian approaches adaptively determine the most relevant coefficients, while providing posterior distributions that quantify uncertainty in the estimates.

The MMSE estimate $\widehat{\mathbf{H}}_{b,\text{MU}} \in \mathbb{C}^{UG_RG_{T,u} \times K}$ can be derived as Eq. \eqref{errorcov_MU} and one can observe that estimating $\widehat{\mathbf{H}}_{b,\textbf{MU}}$ ultimately reduces to the estimation of the corresponding hyperparameter matrix $\mathbf{\Gamma}_{k,\text{MU}}$. Furthermore, the log-likelihood $\text{log}\left[f(\mathbf{Y}_{\text{MU}}(:,k);\mathbf{\Gamma}_{k,\text{MU}})\right]$ of the hyperparameter matrix $\mathbf{\Gamma}_{k,\text{MU}}$ can be expressed as Eq. \eqref{MU_log} where $\mathbf{C}_{w_{y,k}} = \mathbf{C}_w + \mathbf{\Xi}_{\text{MU}}\mathbf{\Gamma}_{k,\text{MU}}\mathbf{\Xi}_{\text{MU}}^H$. However, the maximization of the log-likelihood w.r.t. $\mathbf{\Gamma}_{k,\text{MU}}$ is mathematically intractable. As a remedy, the expectation maximization (EM) \cite{prasad2015joint} algorithm offers an efficient approach for iterative maximization. Furthermore, this algorithm guarantees convergence to the local maximum. 

In preparation for exploiting the EM approach, let the complete information set be defined as $\left\{\mathbf{Y}_{\text{MU}},\mathbf{H}_{b,\text{MU}}\right\}$ and the hyperparameter matrix be represented as $\mathbf{\Gamma}_{\text{MU}}$. Let $\widehat{\mathbf{\Gamma}}_{\text{MU}}^{(j-1)}$ denote the hyperparameter matrix estimate at the $(j-1)$st EM iteration. We now present the detailed update procedure for the hyperparameter estimate $\widehat{\mathbf{\Gamma}}_{\text{MU}}^{(j)}$ in the $j$-th EM iteration. During the E-step, we evaluate the log-likelihood of the complete information set denoted as $\mathcal{L}(\mathbf{\Gamma}_{\text{MU}}|\widehat{\mathbf{\Gamma}}_{\text{MU}}^{(j-1)})$, where
\vspace{-2mm}

\small
\begin{align}
    \mathcal{L}(\mathbf{\Gamma}_{\text{MU}}|\widehat{\mathbf{\Gamma}}_{\text{MU}}^{(j-1)}) & = \mathbb{E}_{\mathbf{H}_{b,\text{MU}}|\mathbf{Y}_{\text{MU}};\widehat{\mathbf{\Gamma}}_{\text{MU}}^{(j-1)}}\left\{\text{log} f (\mathbf{Y}_{\text{MU}},\mathbf{H}_{b,\text{MU}};\mathbf{\Gamma}_{\text{MU}})\right\}.
\end{align}
\normalsize
This can be re-written as
\vspace{-2mm}
\begin{align}
   \mathcal{L}(\mathbf{\Gamma}_{\text{MU}}|\widehat{\mathbf{\Gamma}}_{\text{MU}}^{(j-1)}) = & \mathbb{E}_{\mathbf{H}_{b,\text{MU}}|\mathbf{Y}_{\text{MU}};\widehat{\mathbf{\Gamma}}_{\text{MU}}^{(j-1)}}\big\{\text{log} f (\mathbf{Y}_{\text{MU}}|\mathbf{H}_{b,\text{MU}}) + \notag \\ & \text{log} f (\mathbf{H}_{b,\text{MU}}; \mathbf{\Gamma}_{\text{MU}})\big\}.
\end{align}
\begin{figure*}
\begin{equation}
        \begin{aligned}              
 &\mathbb{E}_{\mathbf{H}_{b,\text{MU}}|\mathbf{Y}_{\text{MU}};\widehat{\mathbf{\Gamma}}_{\text{MU}}^{(j-1)}} \left\{\text{log} f(\mathbf{Y}_{\text{MU}}|\mathbf{H}_{b,\text{MU}})\right\} = \sum_{k=0}^{K-1} \Big( -MN_{RF}^R \text{log}(\pi) - \text{log}[\text{det}(\mathbf{R}_{y,k})]  - \\ & \big(\mathbf{Y}_\text{MU}(:,k) - \mathbf{\Xi}_{\text{MU}}(:,:,k)\mathbf{H}_{b,\text{MU}}(:,k)\big)^H\mathbf{C}_w^{-1}  \big(\mathbf{Y}_{\text{MU}}(:,k) - \mathbf{\Xi}_{\text{MU}}(:,:,k)\mathbf{H}_{b,\text{MU}}(:,k)\big) \Big), \label{esimple}
\end{aligned}
\end{equation}
\hrulefill \vspace{-1.2\baselineskip}
\end{figure*}
The expression within the first expectation operator can be simplified as seen in Eq. \eqref{esimple} given at the top of the next page. One can observe that Eq. \eqref{esimple} is independent of $\mathbf{\Gamma}_{\text{MU}}$, which can be ignored in the subsequent M-step. Thus, the equivalent optimization problem can be formulated as
\begin{align}
    \widehat{\mathbf{\Gamma}}_{\text{MU}}^{(j)} = \mathop{\text{arg max}}_{{\mathbf{\Gamma}}_\text{MU}} \mathbb{E}_{\mathbf{H}_{b,\text{MU}}|\mathbf{Y}_{\text{MU}};\widehat{\mathbf{\Gamma}}_{\text{MU}}^{(j-1)}}\left\{\text{log} f(\mathbf{H}_{b,\text{MU}};\mathbf{\Gamma}_{\text{MU}}) \right\}.
\end{align}
Substituting $f(\mathbf{H}_{b,\text{MU}};\mathbf{\Gamma}_{\text{MU}})$ from Equation \eqref{pri_or}, one can observe that the maximization problem can be separated w.r.t the hyperparameters and one obtains
\begin{equation}
    \begin{aligned}
    \widehat{\mathbf{\Gamma}}_{\text{MU}}^{(j)} = \mathop{\text{arg max}}_{{\mathbf{\Gamma}}_{\text{MU}}} \sum_{i=1}^{U G_R G_{T,u}} & \bigg(-\sum_{k=0}^{K-1} \text{log}(\gamma_{k,i}) - \\ & \sum_{k=0}^{K-1}\frac{1}{\gamma_{k,i}}\mathbb{E}\left\{|\mathbf{H}_{b,\text{MU}}(i,k)|^2\right\}\bigg).
\end{aligned}
\end{equation}
To evaluate the hyperparameter estimate, it is necessary to compute a stationary point of the above cost function. This can be determined by differentiating the cost function above with respect to $\gamma_{k,i}$ and setting the derivative to zero. This leads to the following update equation for the hyperparameter estimate $\widehat{\gamma}_{k,i}^{(j)}$ in the $j$-th EM iteration
\vspace{-2mm}
\begin{align}
    \widehat{\gamma}_{k,i}^{(j)} = \sum_{k=0}^{K-1} \mathbb{E}_{\mathbf{H}_{b,\text{MU}}|\mathbf{Y}_{\text{MU}};\widehat{\mathbf{\Gamma}}_{\text{MU}}^{(j-1)}}\left\{|\mathbf{H}_{b,\text{MU}}(i,k)|^2\right\}, \label{inihyper}
\end{align}
where the conditional expectation in the above equation can be evaluated using the \textit{a posteriori} pdf of $\mathbf{H}_{b,\text{MU}}$ given by $f(\mathbf{H}_{b,\text{MU}}|\mathbf{Y}_{\text{MU}};\widehat{\mathbf{\Gamma}}_{\text{MU}}^{(j-1)}) = \mathcal{CN}(\widehat{\mathbf{H}}_{b,\text{MU}}^{(j)},\mathbf{\Sigma}_{\text{MU}}^{(j)})$. Therefore, the \textit{a posteriori} mean $\widehat{\mathbf{H}}_{b,\text{MU}} \in \mathbb{C}^{UG_RG_{T,u} \times K}$ and \textit{a posteriori} covariance $\boldsymbol{\Sigma}_{\text{MU}} \in \mathbb{C}^{UG_RG_{T,u} \times UG_RG_{T,u} \times K}$ for the MU scenario are derived as Eq. \eqref{MU_mean} and \eqref{MU_sigma} respectively. Note that we adopt a Gaussian prior for the beamspace channel coefficients. This choice is essential for two main reasons. Firstly, the observation model for the received pilots is Gaussian due to AWGN noise, and using a \textit{Gaussian prior ensures conjugacy with the likelihood}. This leads to closed-form expressions for the posterior mean and covariance of the channel coefficients, as given by Eqs. \eqref{MU_mean} and \eqref{MU_sigma} which are directly exploited in the E-step of the EM algorithm. Secondly, by associating each Gaussian prior with a variance hyperparameter that is iteratively refined in the M-step, the proposed framework is able to automatically enforce sparsity. As discussed above, the coefficients with small estimated variances are effectively pruned, while only the dominant beamspace coefficients are retained, thereby yielding accurate sparse channel recovery. Thus, the Gaussian prior is not an arbitrary choice, but a key enabler that provides both computational tractability within EM and the sparsity-inducing behavior that underpins the BGSR algorithm.
\begin{align}
    \widehat{\mathbf{H}}_{b,\text{BGSR}}(:,k) = \mathbf{\Sigma}_{\text{MU}}(:,:,k) &\mathbf{\Xi}_{\text{MU}}^H(:,:,k)\mathbf{C}_w^{-1}\mathbf{Y}_{\text{MU}}(:,k) \notag \\& \: \forall \: 0 \leq k \leq K-1, \label{MU_mean}
\end{align}
\begin{align}
    \mathbf{\Sigma}_{\text{MU}}(:,:,k)  = \big(\mathbf{\Xi}^H_{\text{MU}}(:,:,k) &\mathbf{C}_w^{-1}\mathbf{\Xi}_{\text{MU}}(:,:,k) + \mathbf{\Gamma}_{k,\text{MU}}^{-1}\big)^{-1} \notag \\ & \forall \: 0 \leq k \leq K-1. \label{MU_sigma}
\end{align}
The complexity of the proposed BGSR algorithm is mainly dominated by computing the \textit{a posteriori} covariance matrix as it involves the inverse operation. The inversion in the update has $\mathcal{O}(U^3G_R^3G_{T,u}^3)$ complexity. In order to reduce the complexity, we adopt the \textbf{Woodbury matrix identity} $(\mathbf{A+UCV})^{-1} = \mathbf{A}^{-1} - \mathbf{A}^{-1}\mathbf{U}(\mathbf{C}^{-1} + \mathbf{
V}\mathbf{A}^{-1}\mathbf{U})^{-1}\mathbf{V}\mathbf{A}^{-1}$ and the lemma $(\mathbf{I}+\mathbf{P})^{-1} = \mathbf{I} - (\mathbf{I}+\mathbf{P})^{-1}\mathbf{P}$, where the \textit{a posteriori} covariance and mean can be re-expressed as Eq. \eqref{covari} and \eqref{mean_u} respectively.
\begin{figure*}
    \begin{align}
    \mathbf{\Sigma}_{\text{MU}}(:,:,k) &= \mathbf{\Gamma}_{k,\text{MU}} -  \mathbf{\Gamma}_{k,\text{MU}} \boldsymbol{\Xi}_{\text{MU}}^H(:,:,k)(\mathbf{C}_w^{-1} + \mathbf{\Xi}_{\text{MU}}(:,:,k) \mathbf{\Gamma}_{k,\text{MU}}\mathbf{\Xi}^H_{\text{MU}}(:,:,k))^{-1}\mathbf{\Xi}_{\text{MU}}(:,:,k)\mathbf{\Gamma}_{k,\text{MU}} \: \forall \: 0 \leq k \leq K-1,  \label{covari}\\
    \widehat{\mathbf{H}}_{b,\text{BGSR}}(:,k) &= \mathbf{\Gamma}_{k,\text{MU}} \mathbf{\Xi}_{\text{MU}}^H(:,:,k)\mathbf{C}_w^{-1} \Big(\mathbf{I}_{MN_{\text{RF}}^R} + \mathbf{\Xi}_{\text{MU}}(:,:,k) \mathbf{\Gamma}_{k,\text{MU}} \mathbf{\Xi}_{\text{MU}}^H(:,:,k) \mathbf{C}_w^{-1}\Big)^{-1} \mathbf{Y}(:,k) \: \forall \: 0 \leq k \leq K-1, \label{mean_u}
\end{align}
\hrulefill
\end{figure*}
The inversion required for the new updates of the \textit{a posteriori} covariance matrix has a complexity of $\mathcal{O}\left(U^2G_R^2G_{T,u}^2MN_{RF}^R + M^3(N_{{RF}}^R)^3\right)$, which reduces the overall computational load as $UG_RG_{T,u} \gg MN_{{RF}}^R$. Determining the \textit{a posteriori} PDF from the mean of \eqref{MU_mean} and covariance of \eqref{MU_sigma}, followed by substitution in \eqref{inihyper}, the hyperparameter update finally simplifies to
\vspace{-2mm}
\begin{align}
    \widehat{\boldsymbol{\Gamma}}_{k,\text{MU}}^{(j)} = \text{diag}\left(\mathbf{\Sigma}_{\text{MU}}^{(j)}(:,:,k)+ |\mathbf{H}_{b,\text{MU}}(:,k)|^2\right). \label{hyper-pari}
\end{align}
To refine the hyperparameter estimates further, one can average these parameters across all the subcarriers to glean the processing gain resulting from multiple observations. Thus, the modified hyperparameter update equation is given by
\vspace{-2mm}
\begin{align}
    \widehat{\mathbf{\Gamma}}_{\text{MU}}^{(j)} = \frac{1}{K}\sum_{k=0}^{K-1}\boldsymbol{\Gamma}_{k,\text{MU}}^{(j)}. \label{averaging_hyper}
\end{align}
Note that upon forcing the dictionaries to be subcarrier-dependent due to beam squint, the underlying sparsity pattern of the channel is forced to remain consistent across subcarriers. This allows the hyperparameters to be averaged, since the non-zero coefficient locations do not change with subcarrier. Mathematically, the averaging can be interpreted as approximating the expectation of the hyperparameters across subcarriers, which reduces the variance of the estimates and suppresses the influence of noise or outliers from any individual subcarrier. Thus, the subcarrier-dependent dictionaries capture frequency variation, while hyperparameter averaging improves robustness without contradicting the problem formulation. This trade-off ensures that the estimates are not biased toward any single subcarrier, but instead they reflect the joint statistics across all subcarriers. The MU equivalent sensing matrix $\mathbf{\Xi}_{\text{MU}}$, equivalent received output $\mathbf{Y}_{\text{MU}}$, noise covariance $\mathbf{C}_w$, total subcarriers $K$ and stopping parameters are given by inputs to Algorithm \ref{MU_MSBL_algo} to obtain the group sparse output across all the users. Let the concatenated MU transmit dictionary $\mathbf{A}_{T,\text{MU}} \left(\Theta_T, f_{0:K-1} \right) \in \mathbb{C}^{UN_{T,u} \times UG_{T,u} \times K}$ be formulated as
\vspace{-2mm}
\begin{equation}
    \begin{aligned}
    \mathbf{A}_{T,\text{MU}} & \left(\Theta_T, f_{0:K-1} \right) =  \text{blkdiag}\Big(\mathbf{A}_{T,\text{1}}(\Theta_{T,1}, f_{0:K-1}), \\ & \mathbf{A}_{T,\text{2}}(\Theta_{T,2}, f_{0:K-1}),  \cdots, \mathbf{A}_{T,U}(\Theta_{T,U}, f_{0:K-1})\Big). \label{MU_trans}
\end{aligned}
\end{equation}
The estimated channel obtained through the BGSR framework can now be determined as
\vspace{-5mm}

\small
\begin{align}
    \widehat{\mathbf{H}}_{\text{BGSR}} = \mathbf{A}_{R}(\Phi_R, f_{0:K-1}) \text{vec}^{-1}(\widehat{\mathbf{H}}_{b,\text{BGSR}}) \mathbf{A}_{T,\text{MU}}^H(\Theta_T, f_{0:K-1}) \label{MU_gsomp}.
\end{align}
\normalsize
\begin{algorithm}[t]
\DontPrintSemicolon 
\KwIn{$\mathbf{\Xi}_{\text{MU}}$, $\mathbf{Y}_{\text{MU}}$, $\mathbf{C}_w$, $K$, $\mathbf{A}_R(\Phi_R, f_{0:K-1})$ and $\mathbf{A}_{\mathit{T}, \text{MU}}(\Theta_T, f_{0:K-1})$, $\epsilon$ and $\mathit{K}_{\text{max}}$}
\KwOut{$\widehat{\mathbf{H}}_{\textrm{BGSR}}= \mathbf{A}_R(\Phi_R, f_{0:K-1})\textrm{vec}^{-1}(\widehat{\mathbf{H}}_{b,\text{BGSR}})\mathbf{A}_{T,\text{MU}}^H(\Theta_T, f_{0:K-1})$}
\textbf{Initialization:} $\widehat{\mathbf{\Gamma}}^{(1)}_{\text{MU}} = \mathbf{I}_{U\mathit{G}_R\mathit{G}_{T,u}}$, $\widehat{\mathbf{\Gamma}}^{(0)}_{\text{MU}} = \mathbf{0}_{UG_RG_{T,u}}$ and counter $\mathit{j} = \textrm{1}$
 
\While{$\big(\parallel \widehat{\mathbf{\Gamma}}_{\mathrm{MU}}^{\left(\mathit{j}\right)} - \widehat{\mathbf{\Gamma}}_{\mathrm{MU}}^{\left(\mathit{j}-1\right)}\parallel_{\mathit{F}}^2 \ > \epsilon \ \&\& \ \mathit{j} < \mathit{K}_{\mathrm{max}}\big)$}
{
 
\textbf{E-Step:} Compute the \textit{a posteriori} covariance and mean as

\For{$\mathit{k} = 0,\cdots, K-1$}
 {
 Compute the covariance using Eq. \eqref{covari}\\
 Compute the mean using Eq. \eqref{mean_u}
}
 
 \textbf{M-Step:} Update the estimates of the hyperparameters as

\For{$\mathit{k} = 0,\cdots, K-1$}
 {
Update the hyperparameters using Eq. \eqref{hyper-pari}.
}

Averaging hyperparameter using Eq. \eqref{averaging_hyper}

$j \leftarrow j+1$ 
 
}

\textbf{return:~~}{$\widehat{\mathbf{H}}_{\mathit{b},\text{BGSR}}$}
\caption{BGSR based sparse channel estimation for dual-wideband MU THz MIMO systems}
\label{MU_MSBL_algo}
\end{algorithm}
A significant implication of the proposed BGSR framework is the sparsity of the distribution across distinct dictionaries. This implies that when we use a common dictionary across all subcarriers, the sparse locations will vary for each subcarrier. However, when we employ a dictionary corresponding to each subcarrier, it fixes the sparse locations, resulting in group sparsity in the channel. In essence, the sparse locations remain consistent across different subcarrier dictionaries, despite each subcarrier having its unique dictionary matrix. This characteristic further enhances the robustness of the algorithm, which makes it novel and can be used in practical scenarios. A popular choice for embedding the non-linearity caused by low-resolution ADCs includes Bayesian inference \cite{mo2017channel}, which corresponds to a discrete likelihood. However, in a dual-wideband THz system associated with a large number of antennas and subcarriers, this discrete likelihood must be recalculated iteratively for each subcarrier, leading to substantial memory usage and computational overhead. As a result, maintaining a Bayesian inference loop under these conditions becomes prohibitively demanding, making closed-form iterative updates elusive, unless additional approximations or dimensionality reduction techniques are introduced. The next subsection briefly describes the GSMP-based sparse channel estimation method, originally presented in \cite{dovelos2021channel} for THz communication and later extended for the MU-MIMO system model with low-resolution ADCs exploiting cross-subcarrier sparsity.
\subsection{Group Structure Matching Pursuit based sparse channel estimation}
In order to effectively compare our proposed MU BGSR approach, this treatise also introduces a MU group structure matching pursuit (GSMP)-based channel learning technique, which is described in the following section. The concatenated group sparse based channel model is detailed in Section \ref{G-MSB}. Moreover, the MU-equivalent sensing matrix $\mathbf{\Xi}_{\text{MU}}$, the received output $\mathbf{Y}_{\text{MU}}$, and the stopping parameter are fed into Algorithm-\ref{MU_gsomp_algo} to obtain the estimated sparse channel for all the users. The \textit{greedy-iterative} procedure consists of three distinct phases which includes \textit{Identification} (Steps $5,6,7$), \textit{Augmentation} (Steps $9$ and $10$) and the \textit{Residual Updation} phase. Step $5$ performs summed correlation across all subcarriers, rather than independently selecting the strongest correlation for each subcarrier, as it is done in standard OMP \cite{cheng2016matrix}. Step $6$ then updates the residual in a grouped manner. While the existing MMV-based OMP approach \cite{gonzalez2018channel} also processes the subcarriers jointly, they use a common dictionary for computing a single correlation metric aggregated both across all measurement vectors in each iteration. This typically selects one index at a time based on the aggregated correlation. By contrast, our GSMP variant in Algorithm-$2$ performs a more refined, group-based selection (as shown in Step $5$ and $6$), where multiple subcarrier indices are chosen to refine the group structure in each iteration. Additionally, Step $9$ augments the empty matrix with columns obtained from the index set of Step $7$, while Step $10$ computes the intermediate LS solution from the matrix $\mathbf{\Xi}_{\text{MU}}^{\mathcal{A}}$. Finally, Step $11$ updates the residual vector by subtracting the LS solution obtained from the original output vector, whereas Step $12$ averages the residual vector at each iteration. The estimated channel can be reconstructed using Eq. \eqref{MU_gsomp} corresponding to all the users. Note that the GSMP algorithm is introduced in this work to demonstrate the \textit{trade-off} between computational complexity and estimation accuracy in comparison to the BGSR framework and  it is also one of the pair of algorithms proposed.
\begin{algorithm}[t]
\DontPrintSemicolon 
\KwIn{$\mathbf{\Xi}_{\text{MU}}$, $\mathbf{Y}_{\text{MU}}$, $\mathbf{A}_R(\Phi_R,f_{0:K-1})$ and $\mathbf{A}_{T,\text{MU}}(\Theta_T,f_{0:K-1})$, $\epsilon_0$}
\KwOut{$\widehat{\mathbf{H}}_{\text{GSMP}} = \mathbf{A}_R(\Phi_R, f_{0:K-1}) \text{vec}^{-1}(\widehat{\mathbf{H}}_{\mathit{b},\text{GSMP}}) \mathbf{A}_{T,\text{MU}}^H(\Theta_T, f_{0:K-1})$}
\textbf{Initialization:} Index set $\mathcal{A} = [\ ]$, Empty matrix $\mathbf{\Xi}_{\text{MU}}^{\mathcal{A}} = [\ ]$, $\widehat{\mathbf{H}}_{\text{LS},\text{MU}} = [\ ]$, Residue vector $\mathbf{T}_0 = \mathbf{0}_{\mathit{M}\mathit{N}_{\mathit{RF}}^R\times K}, \mathbf{T}_1 = \mathbf{Y}_{\text{MU}}$, Estimated beamspace channel $\widehat{\mathbf{H}}_{b,\text{GSMP}} = \mathbf{0}_{U G_R G_{T,u} \times K}$, counter $j = 1$ 
 
\While{$\left( \left\vert\parallel \mathbf{T}_{j-1} \parallel_{F}^2 - \parallel \mathbf{T}_{j} \parallel_{F}^2 \right\vert\ \geq \ \epsilon_0 \right)$}
{

$\mathbf{Z}_{\text{MU}} = [\ ]$

 \For{$k = 0,1,\ldots,K-1$}
 {

$\mathbf{Z}_{\text{MU}} = \mathbf{Z}_{\text{MU}} \cup \left(\mathbf{\Xi}_{\text{MU}}^H(:,:,k)\mathbf{T}_j(:,k)\right)$
}

$\xi = \textrm{arg\ max} \underset{k}{\sum} |\mathbf{Z}_{\text{MU}}(:,k)|^2$
 
$\mathcal{A} = \mathcal{A} \cup \xi$

 \For{$k = 0,1,\ldots,K-1$}
 {
$\mathbf{\Xi}_{\text{MU}}^{\mathcal{A}} = \mathbf{\Xi}_{\text{MU}}(:,\mathcal{A},k)$
 
$\widehat{\mathbf{H}}_{\text{LS},\text{MU}}(:,\mathit{k}) = (\mathbf{\Xi}_{\text{MU}}^{\mathcal{A}})^{\dagger}\mathbf{Y}_{\text{MU}}(:,k)$

$\mathbf{T}_{\mathit{j}}(:,k) = \mathbf{Y}_{\text{MU}}(:,k) - \mathbf{\Xi}_{\text{MU}}^{\mathcal{A}}\widehat{\mathbf{H}}_{\textrm{LS},\text{MU}}(:,\mathit{k})$
}

$\mathbf{T}_j = \frac{\mathbf{T}_j}{K}$

$j \leftarrow j+1$
 
}

$\widehat{\mathbf{H}}_{b,\text{GSMP}}(\mathcal{A},0:K-1) = \widehat{\mathbf{H}}_{\textrm{LS},\text{MU}}$

\textbf{return:~~}{$\widehat{\mathbf{H}}_{\mathit{b},\text{GSMP}}$}
\caption{GSMP based sparse channel estimation for dual-wideband MU THz MIMO systems}
\label{MU_gsomp_algo}
\end{algorithm}

\textbf{Remark 1:} In order to relax the grid-based dictionary constraint, we also adopt the continuous basis pursuit \cite{ekanadham2011recovery} based dictionary, which is approximated by the first order Taylor series interpolation of
\vspace{-4mm}

\small
\begin{align}
    \Tilde{\mathbf{a}}(\phi,f_k) = \Tilde{\mathbf{a}}(\phi_r,f_k) + (\phi - \phi_r) \frac{\partial \Tilde{\mathbf{a}}(\phi,f_k)}{\partial \phi} \bigg|_{\phi_r} + \mathcal{O}((\phi-\phi_r)^2),
\end{align}
\normalsize
where $\phi$ represents the continuous AoA/ AoD, while $\phi_r = 2 \pi \frac{(r-1)}{G_R}$ represents the grid point having the minimal distance from $\phi$. Therefore, the new dictionary will contain the grid points from the quantized array factor along with its derivatives, which can be further formulated as Eq. \eqref{deriva}
\begin{align}
    \boldsymbol{\mathfrak{T}}_R(\phi_r,f_k) = [\Tilde{\mathbf{a}}(\phi_1,f_k),\cdots,\Tilde{\mathbf{a}}(\phi_{G_R},f_k), \notag \\
    \Tilde{\mathbf{b}}(\phi_1,f_k),\cdots,\Tilde{\mathbf{b}}(\phi_{G_R},f_k)], \label{deriva}
\end{align}
where $\Tilde{\mathbf{b}}(\phi_r,f_k) = \frac{\partial \Tilde{\mathbf{a}}(\phi,f_k)}{\partial \phi}\Big|_{\phi_r}$. Furthermore, to incorporate the offset angle wrt the grid, an interpolator is defined as
\begin{align}
    \boldsymbol{\mathfrak{t}}_R = \Big[\underbrace{1,\cdots,1}_{G_R},\underbrace{\triangle \phi,\cdots,\triangle \phi}_{G_R}\Big],
\end{align}
where $|\triangle \phi| \leq \frac{\pi}{G_R}$. Finally, the Taylor series based off-grid dictionary (TBoD) is given by Eq. \eqref{TBO-D}.
\begin{figure*}
    \begin{align}
     \Tilde{\mathbf{A}}_R(\Phi_R,f_k) &= \boldsymbol{\mathfrak{T}}_R(\phi_r,f_k) \boldsymbol{\mathfrak{t}}_R  = [\Tilde{\mathbf{a}}_R(\phi_1,f_k),\cdots,\Tilde{\mathbf{a}}_R(\phi_{G_R},f_k),\triangle \phi \Tilde{\mathbf{b}}_R(\phi_1,f_k),\cdots, \triangle \phi \Tilde{\mathbf{b}}_R(\phi_{G_R},f_k)]. \label{TBO-D}\\
     \tilde{\mathbf{A}}_{T,u}(\Theta_{T,u},f_k) &= [\Tilde{\mathbf{a}}_{T,u}(\theta_1,f_k),\cdots,\Tilde{\mathbf{a}}_{T,u}(\theta_{G_{T,u}},f_k),\triangle \theta \Tilde{\mathbf{b}}_{T,u}(\theta_1,f_k),\cdots, \triangle \theta \Tilde{\mathbf{b}}_{T,u}(\theta_{G_{T,u}},f_k)]. \label{trans_TBoD}
\end{align} \vspace{-1.2 \baselineskip}
\hrulefill
\end{figure*}
Note that, a similar TBoD matrix is formed at the transmitter, which is given by Eq. \eqref{trans_TBoD}. The next section presents the computational complexity of the proposed algorithm along with that of GSMP algorithm.
\subsection{Computational Complexity}
This subsection derives the computational complexities of the BGSR and GSMP based THz channel learning techniques. Table-\ref{GSMP_complexity} list the per-subcarrier and per-iteration complexity of the GSMP-framework along with various key computational steps. The overall computational complexity of the GSMP-framework is on the order of $\left(KUG_RG_{T,u}MN_{\text{RF}}^R+K(\frac{2}{3}j^3+j^2MN_{\text{RF}}^R+3jMN_{\text{RF}}^R)\right)$, where $j$ represents the current iteration. The worst-case complexity order is $\mathcal{O}(K(M^3N_{\text{RF}}^R)^3),$ which is attributed to the necessity of an intermediate LS estimate at each iteration. Table-\ref{BGSR_complexity} lists the per-subcarrier and per-iteration complexity of the BGSR-framework along with various key computational steps. The worst-case complexity of the BGSR technique for each subcarrier is on the order of $\mathcal{O}\big(2(MN_{\text{RF}}^R)^3+2UG_RG_{T,u}(MN_{\text{RF}}^R)^2+3UG_RG_{T,u}MN_{\text{RF}}^R+(UG_RG_{T,u})^2MN_{\text{RF}}^R\big),$ which is dominated by the matrix-inversion of $\mathbf{C}_w.$ Moreover, as discussed above, $UG_RG_{T,u} \gg MN_{\text{RF}}^R$; therefore, the approximated complexity of the BGSR algorithm for each subcarrier is $\mathcal{O}(U^2G_R^2G_{T,u}^2MN_{\text{RF}}^R + M^3(N_{\text{RF}}^R)^3)$. The worst-case computational complexity for all the subcarriers is $\mathcal{O}\big(K(U^2G_R^2G_{T,u}^2MN_{\text{RF}}^R + M^3(N_{\text{RF}}^R)^3)\big),$ implying that the computational complexity increases linearly with the number of subcarriers and also higher than GSMP framework. However, as shown in Section-\ref{Simulation_results}, BGSR indicates superior performance. Hence, there is a clear trade-off between computational complexity and estimation performance. The BCRB characterizing the estimation of the MU channel is derived in detail next.
\begin{table*}[]
    \centering
    \vspace{-2mm}
    \caption{Computation complexity of GSMP scheme, per-subcarrier in the $j$-th iteration}
    \label{GSMP_complexity}
    \resizebox{1\textwidth}{!}{ 
    \begin{tabular}{|c|c|c|}
        \hline
        \textbf{Operation} &  \textbf{Complex multiplications} & \textbf{Complex additions} \\ \hline
        $\boldsymbol{\Xi}_{\text{MU}}^H(:,:,k)\mathbf{T}(:,k)$ & $UG_RG_{T,u}MN_{\text{RF}}^R$ & $UG_RG_{T,u}(MN_{\text{RF}}^R-1)$ \\ \hline
        $\widehat{\mathbf{H}}_{\text{LS,MU}}(:,k) = \left(\boldsymbol{\Xi}_{\text{MU}}^\mathcal{A}\right)^\dagger\mathbf{Y}_{\text{MU}}(:,k)$ & $\frac{2}{3}j^3 + j^2MN_{\text{RF}}^R+2jMN_{\text{RF}}^R$ & $\frac{2}{3}j^3+j^2(MN_{\text{RF}}^R-1)+j(MN_{\text{RF}}-1)$ \\ \hline
        $\mathbf{Y}_{\text{MU}}(:,k)-\boldsymbol{\Xi}_{\text{MU}}^\mathcal{A}\widehat{\mathbf{H}}_{\text{LS,MU}}(:,k)$ & $jMN_{\text{RF}}^R$ & $jMN_{\text{RF}}^R$ \\ \hline
    \end{tabular}} \vspace{-0.5 \baselineskip}\\
\end{table*}
\begin{table*}[]
    \centering
    \vspace{-2mm}
    \caption{Computation complexity of BGSR framework, per-subcarrier, per-EM iteration}
    \label{BGSR_complexity}\resizebox{1\textwidth}{!}{
    \begin{tabular}{|c|c|c|}
        \hline
        \textbf{Operation} &  \textbf{Complex multiplication} & \textbf{Complex additions} \\ \hline
        $\mathbf{\Gamma}_{k,\text{MU}}\boldsymbol{\Xi}_{\text{MU}}^H(:,:,k)$ & $UG_RG_{T,u}MN_{\text{RF}}^R$ & $0$ \\ \hline
        $\mathbf{C}_w^{-1}$ & $\frac{(MN_{\text{RF}}^R)^3}{2}+\frac{3(MN_{\text{RF}}^R)^2}{2}$ & $\frac{(MN_{\text{RF}}^R)^3}{2}-\frac{3(MN_{\text{RF}}^R)^2}{2}$ \\ \hline
        $\boldsymbol{\Xi}_{\text{MU}}(:,:,k)\mathbf{\Gamma}_{k,\text{MU}}\boldsymbol{\Xi}_{\text{MU}}^H(:,:,k)$ & \makecell{$(MN_{\text{RF}}^R)^2UG_RG_{T,u}$ \\ + $MN_{\text{RF}}^RUG_RG_{T,u}$} & \makecell{$(MN_{\text{RF}}^R)^2(UG_RG_{T,u}-1)$ \\ + $(MN_{\text{RF}}^R)(UG_RG_{T,u}-1)$} \\ \hline
        \makecell{$\big(\mathbf{C}_w^{-1}+\boldsymbol{\Xi}_{\text{MU}}(:,:,k)$\\$\mathbf{\Gamma}_{k,\text{MU}}\boldsymbol{\Xi}_{\text{MU}}^H(:,:,k)\big)^{-1}$} & \makecell{$\frac{(MN_{\text{RF}}^R)^3}{2}+\frac{3UG_RG_{T,u}(MN_{\text{RF}}^R)^2}{2}+$ \\ $UG_RG_{T,u}MN_{\text{RF}}^R$} & \makecell{$\frac{(MN_{\text{RF}}^R)^3}{2}-\frac{3UG_RG_{T,u}(MN_{\text{RF}}^R)^2}{2}-$ \\ $UG_RG_{T,u}MN_{\text{RF}}^R$} \\ \hline
    \end{tabular}} \vspace{-1 \baselineskip}
\end{table*}
\subsection{MU Bayesian Cram{\'e}r-Rao bound}
Consider the parameterized Gaussian prior associated with the beamspace channel matrix $\mathbf{H}_{b,\text{MU}}$ as
\begin{align}    f(\mathbf{H}_{b,\text{MU}};\mathbf{\Gamma}_{\text{MU}}) = \prod_{k=0}^{K-1} \frac{1}{(\pi)^{UG_RG_{T,u}} \text{det}(\mathbf{\Gamma}_{k,\text{MU}})} \notag \\ \text{exp}\big(-\mathbf{H}_{b,\text{MU}}^H(:,k)\mathbf{\Gamma}_{k,\text{MU}}^{-1}\mathbf{H}_{b,\text{MU}}(:,k)\big). \label{log_likeli}
\end{align}
The log-likelihood of the beamspace channel $\mathbf{H}_{b,\text{MU}}$ is given as 
\begin{align}
 \mathcal{L}(f(\mathbf{H}_{b,\text{MU}};\mathbf{\Gamma}_{\text{MU}})) = \sum_{k=0}^{K-1} \Big( -UG_RG_{T,u} \text{log}(\pi) - \notag \\ \text{log}[\text{det}(\mathbf{\Gamma}_{k,\text{MU}})] - \mathbf{H}_{b,\text{MU}}^H(:,k)\mathbf{\Gamma}_{k,\text{MU}}^{-1}\mathbf{H}_{b,\text{MU}}(:,k) \Big). \label{log_likelih}
\end{align}
The conditional PDF of the pilot output vector $\mathbf{Y}_{\text{MU}}$, is represented by the complex normal distribution $\prod_{k=0}^{K-1} \mathcal{CN}(\mathbf{\Xi}_{\text{MU}}(:,:,k)\mathbf{H}_{b,\text{MU}}(:,k), \mathbf{C}_w)$. Therefore, the conditional PDF is given by Equation \eqref{MU_cond_pd}, where $\Upsilon = \frac{1}{(\pi)^{MN_{RF}^R}\text{det}(\mathbf{C}_w)}$. Taking the logarithm on both sides of Equation \eqref{MU_cond_pd} yields Equation \eqref{Loglikefn}. Let $\mathbf{I}_{\text{MU}} \in \mathbb{C}^{UG_RG_{T,u} \times UG_RG_{T,u}}$ denote the Bayesian Fisher information matrix (FIM), formulated as
\begin{align}
    \mathbf{I}_{\text{MU}} = \mathbf{I}_{y,\text{MU}} + \mathbf{I}_{b,\text{MU}},
\end{align}
where $\mathbf{I}_{y,\text{MU}} \in \mathbb{C}^{UG_RG_{T,u} \times UG_RG_{T,u}}$ denotes the FIM associated with the combined received signal $\mathbf{Y}_{\text{MU}}$, while $\mathbf{I}_{b,\text{MU}} \in \mathbb{C}^{UG_RG_{T,u} \times UG_RG_{T,u}}$ quantifies the FIM arising from the \textit{a priori} knowledge about the beamspace CSI matrix $\mathbf{H}_{b,\text{MU}}$. The following equations represent these quantities mathematically
\vspace{-4mm}
\begin{align}
    \mathbf{I}_{y,\text{MU}} = - \mathbb{E}_{\mathbf{Y}_{\text{MU}},\mathbf{H}_{b,\text{MU}}}\left\{\frac{\partial^2 \mathcal{L}(\mathbf{Y}_{\text{MU}}|\mathbf{H}_{b,\text{MU}})}{\partial\mathbf{H}_{b,\text{MU}} \, \partial\mathbf{H}_{b,\text{MU}}^H}  \right\},
\end{align}
\vspace{-4mm}
\begin{align}
    \mathbf{I}_{b,\text{MU}} = -\mathbb{E}_{\mathbf{H}_{b,\text{MU}}}\left\{\frac{\partial^2 \mathcal{L}(\mathbf{H}_{b,\text{MU}};\mathbf{\Gamma}_{\text{MU}})}{\partial\mathbf{H}_{b,\text{MU}} \, \partial\mathbf{H}_{b,\text{MU}}^H}\right\}.
\end{align}
\begin{figure*}
 \begin{align}
     f(\mathbf{Y}_{\text{MU}}|\mathbf{H}_{b,\text{MU}}) = \prod_{k=0}^{K-1} \Upsilon \text{exp}\Big(- \big(\mathbf{Y}_{\text{MU}}(:,k) - \mathbf{\Xi}_{\text{MU}}(:,:,k)\mathbf{H}_{b,\text{MU}}(:,k) \big)^H \mathbf{C}_w^{-1} \big(\mathbf{Y}_{\text{MU}}(:,k) - \mathbf{\Xi}_{\text{MU}}(:,:,k)\mathbf{H}_{b,\text{MU}}(:,k)\big) \Big), \label{MU_cond_pd}
 \end{align}
 \vspace{-5mm}
 \begin{align}
     \mathcal{L}(\mathbf{Y}_{\text{MU}}| \mathbf{H}_{b,\text{MU}}) = \sum_{k=0}^{K-1} \Big(-MN_{RF}^R \text{log}(\pi) - \text{log}[\text{det}(\mathbf{C}_w)] - & \big(\mathbf{Y}_{\text{MU}}(:,k) - \mathbf{\Xi}_{\text{MU}}(:,:,k)\mathbf{H}_{b,\text{MU}}(:,k) \big)^H \mathbf{C}_w^{-1} \notag \\ & \big(\mathbf{Y}_{\text{MU}}(:,k) - \mathbf{\Xi}_{\text{MU}}(:,:,k)\mathbf{H}_{b,\text{MU}}(:,k)\big)\Big). \label{Loglikefn}
 \end{align}
\hrulefill \vspace{-1 \baselineskip}
\end{figure*}
Using the conditional PDF as derived in Equation \eqref{Loglikefn}, the FIM $\mathbf{I}_{y,\text{MU}}$ can be modeled as
\begin{align}
    \mathbf{I}_{y,\text{MU}} = \sum_{k=0}^{K-1} \mathbf{\Xi}_{\text{MU}}^H(:,:,k) \mathbf{C}_w^{-1} \mathbf{\Xi}_{\text{MU}}(:,:,k). \label{MU_FIM_yk}
\end{align}
In a similar vein, the FIM corresponding to the $\mathbf{H}_{b,\text{MU}}$ can be derived as
\vspace{-4mm}
\begin{align}
    \mathbf{I}_{b,\text{MU}} = \widehat{\mathbf{\Gamma}}_{\text{MU}}^{-1}.\label{MU_FIM_hb}
\end{align}
The complete Bayesian FIM is now obtained by combining Equation \eqref{MU_FIM_yk} and \eqref{MU_FIM_hb}, which is formulated as
\begin{align}
    \mathbf{I}_{\text{MU}} = \sum_{k=0}^{K-1} \big(\mathbf{\Xi}_{\text{MU}}^H(:,:,k) \mathbf{C}_w^{-1} \mathbf{\Xi}_{\text{MU}}(:,:,k)\big) + \widehat{\mathbf{\Gamma}}_{\text{MU}}^{-1}.
\end{align}
We further define the equivalent sparsifying dictionary across all subcarriers as $\mathbf{\Delta}_{\text{MU}} = \left[\mathbf{\Psi}_{\text{MU}}[0] \: \mathbf{\Psi}_{\text{MU}}[1] \:  \cdots \: \mathbf{\Psi}_{\text{MU}}[K-1] \right] \in \mathbb{C}^{U N_R N_{T,u} \times U G_R G_{T,u} \times K}$. Finally, the BCRB for the MSE of the estimated CSI is expressed as
\begin{align}
    \text{MSE}(\widehat{\mathbf{H}}_{\text{MU}}) \geq \text{Tr}\left(\mathbf{\Delta}_{\text{MU},0:K-1}\mathbf{I}_{\text{MU}}^{-1}\mathbf{\Delta}_{\text{MU},0:K-1}^H\right).
\end{align}
\vspace{-4mm}
\section{Simulation Results} \label{Simulation_results}
This section presents our results characterizing the performance of the proposed channel learning technique, namely the BGSR designed for MU THz hybrid MIMO systems. Table-\ref{simulation_para} provides the numerical parameters considered for the simulation study. Furthermore, the root raised cosine and rectangular pulse shapes denoted by RRC-PSF and Rect-PSF, respectively, are considered for performance evaluation. Note that, Rect-PSF represents the dual-wideband channel when a rectangular filter is used instead of Root Raised Cosine filter in Equation \eqref{pulse_shaping}.
\begin{table}[t]
\centering
\caption{Simulation parameters for the dual-wideband THz hybrid MIMO system}
\label{simulation_para}
\resizebox{0.42\textwidth}{!}{%
\begin{tabular}{|l|c|}\hline
\textbf{Parameter} & \textbf{Value} \\ \hline \hline
$\#$ of TAs ($N_T$)  & $4$ \\\hline
$\#$ of RAs ($N_R$)  & $48$ \\ \hline
$\#$ of TAs RF chains ($N_{RF}^T$)  & $2$ \\ \hline
$\#$ of RAs RF chains ($N_{RF}^R$)  & $8$ \\ \hline
$\#$ of subcarriers ($K$) &  $64$ \\ \hline
$\#$ of pilot blocks ($M$)   & $20$ \\ \hline
$\#$ of data vectors ($N_d$)  & $100$ \\ \hline
$\#$ of Users ($U$)  & $3$ \\ \hline
$\#$ of NLoS components ($N_{\text{NLoS}}$) &  $3$ \\ \hline
$\#$ of delay taps ($L$) &  $3$ \\ \hline
Transmit Antenna Gain ($\mathfrak{B}_T$) &  $31 \: \text{dBi}$ \\ \hline
Receive Antenna Gain ($\mathfrak{B}_R$) &  $31 \: \text{dBi}$ \\ \hline
Transmit angular grid size ($G_{T,u}$) & $8$ \\ \hline
Receive angular grid size ($G_R$)  & $96$ \\ \hline
Operating frequency ($f_c$) & $0.65 \: \text{THz}$ \\\hline
Transmission distance ($d$) & $15 \text{m}$ \\\hline
Bandwidth ($B$) & $5 \: \text{GHz}$ \\\hline
Angle quantization parameter ($N_Q$) & $4$ \\\hline
ADC Resolution & $3$-\text{bit} \\\hline
Constellation & $8$-\text{PSK} \\\hline
Roll-off factor for RRC-PSF & $0.80$ \\\hline
Upsampling factor & $20$ \\ \hline
GSMP Threshold ($\epsilon_0$) & $2$ \\ \hline
BGSR Thresholds ($\epsilon, K_{\text{max}}$) & $1, 20$ \\
\hline
\end{tabular}%
}
\end{table}
\begin{table}[t]
\centering
\caption{List of materials used for simulation environment \cite{piesiewicz2007scattering},\cite{piesiewicz2007properties}} \label{materials_para}
\resizebox{0.42\textwidth}{!}{%
\begin{tabular}{|l|c|c|c|}\hline
\textbf{Material Type} & $\sigma_r (\mathrm{in \: mm})$ & $\varkappa (\mathrm{in \: cm^{-1}})$ & $\eta$ \\\hline
Polycarbonate (PC) & $0$ & $23$ & $1.52$  \\\hline
Polystyrene (PS) & $0.002$ & $2$& $1.6$ \\\hline
Polyvinyl chloride (PVC) & $0.028$ & $19$ & $1.68$  \\\hline
Plaster s1 & $0.05$ & $10$ & $2$ \\\hline
Gypsum plaster & $0.13$ & $38$ & $1.4$  \\\hline
Plaster s2 & $0.15$ & $10$ & $2$  \\\hline
\end{tabular}%
}
\vspace{-0.8\baselineskip}
\end{table}
\begin{figure*}
	\centering
\subfloat[]{\includegraphics[scale=0.31]{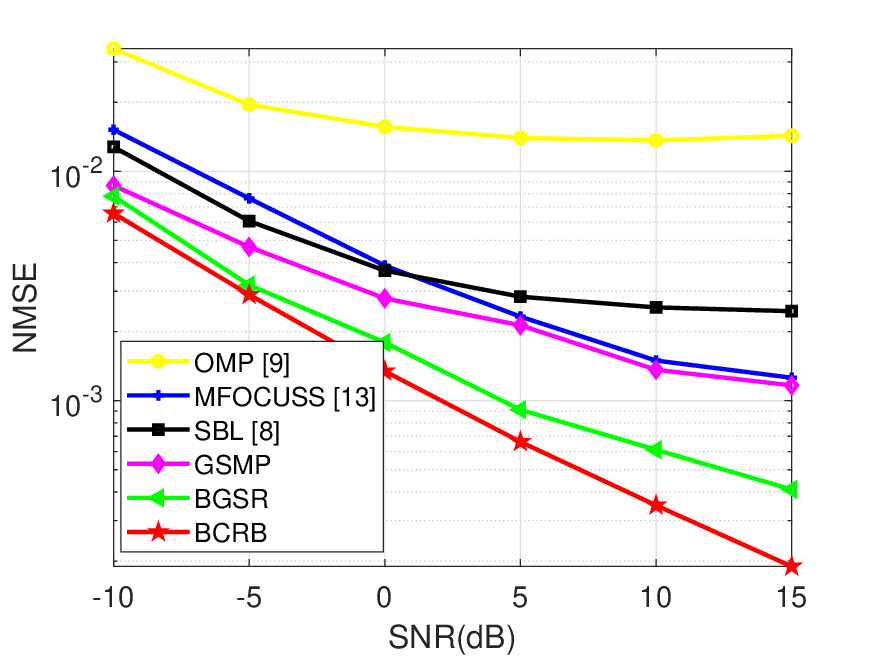}}
	\hfil
	\hspace{-5pt}\subfloat[]{\includegraphics[scale=0.31]{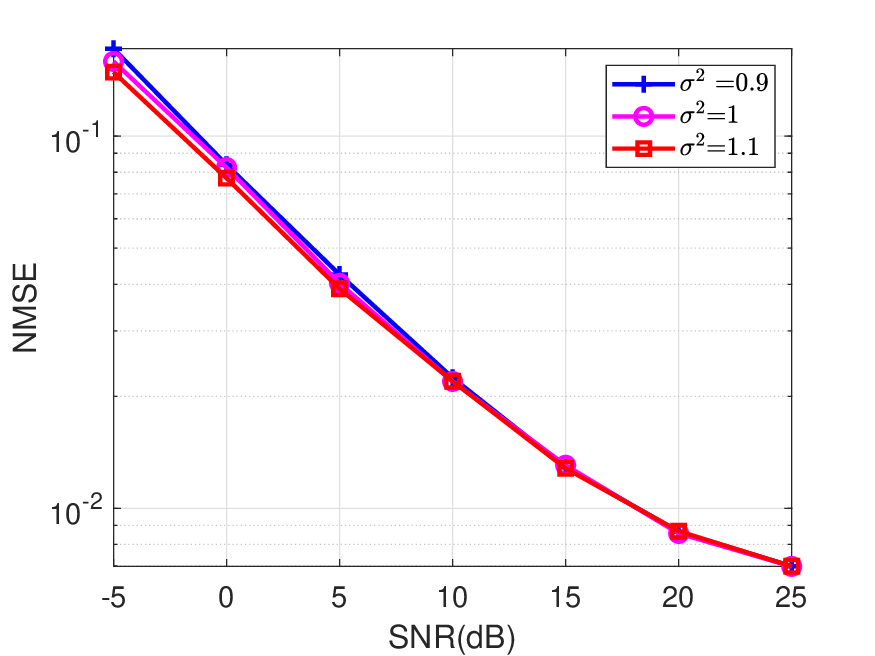}}
            \hfil
	\hspace{-5pt}\subfloat[]{\includegraphics[scale=0.31]{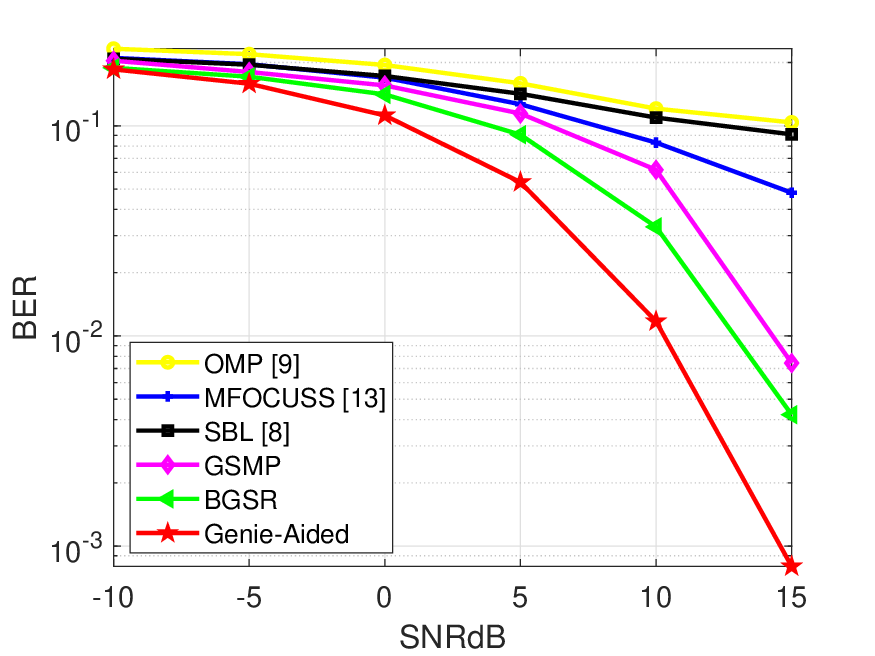}}
        \hfil
	\hspace{-5pt}\subfloat[]{\includegraphics[scale=0.31]{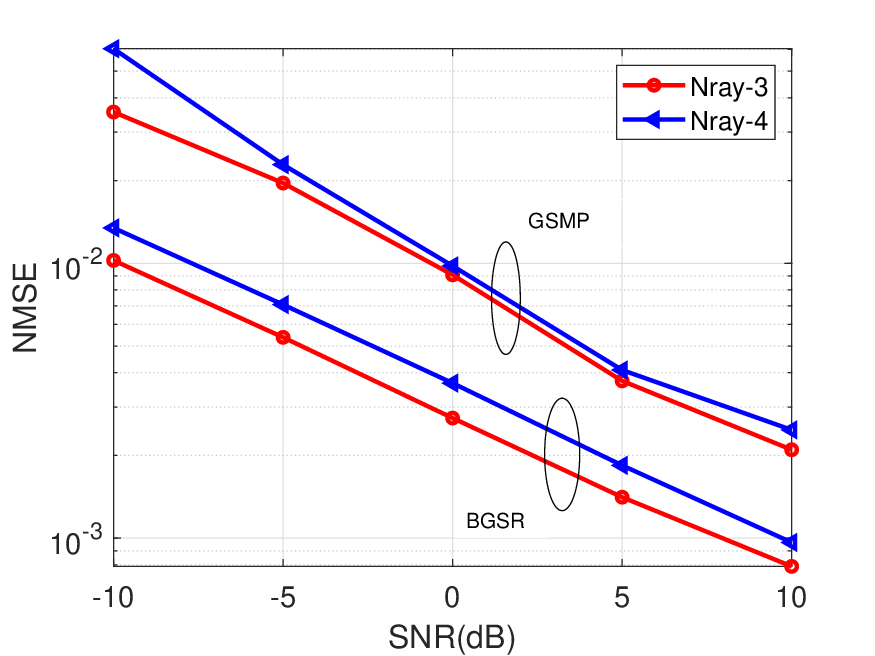}}
 \vspace{-2mm}
	\caption{ $ \left(a\right) $ NMSE vs SNR performance comparison of the proposed GSMP and BGSR based sparse channel estimation techniques with the existing state-of-the-art techniques $ \left(b\right) $  NMSE vs SNR comparison under varying noise variance $ \left(c\right) $ BER vs SNR performance comparison of the proposed GSMP and BGSR based sparse channel estimation techniques with the existing state-of-the-art techniques $ \left(d\right) $ Effect of increasing diffused rays for MU scenarios.}\label{MU_BCRLB}
\end{figure*}
The quantity $\mathfrak{B}_T$ in Table-\ref{simulation_para} represents the combined transmit gain for all the users, i.e., $\sum_{u=1}^U \mathfrak{B}_{T,u} = \mathfrak{B}_T$. To generate the amplitudes of the LoS and NLoS components, for both the dual wideband formulations, we utilize Equations \eqref{LoS_component} and \eqref{NLoS_component} respectively. The phase shifts corresponding to \eqref{LoS_component} and \eqref{NLoS_component} are generated as independent samples of a uniformly distributed random variable over the interval $(-\pi,\pi]$. Both the transmit and receive antennas maintain inter-antenna spacings of $d_t = d_r = \frac{\lambda}{2}$. The MA coefficient appearing in Equations \eqref{LoS_component} and \eqref{NLoS_component} is calculated using the HITRAN database \cite{jornet2011channel}. We consider an indoor office scenario with a molecular composition of $20.9$\% oxygen, $78.1$\% nitrogen and $1$\% water vapour, where the temperature and pressure are set to $298$ K and $1$ atm, respectively. We consider $6$ scatters that can be present in a general office scenario, which are detailed in Table-\ref{materials_para}. Furthermore, the data symbols are randomly generated with an average power of unity. Therefore, the SNR in decibels (dB) is calculated as SNR(dB) = $10 \log_{10} \left(\frac{1}{\sigma^2}\right)$. The receiving and transmitting array response matrices are generated using \eqref{receive_array} and \eqref{MU_trans} respectively, and the grid-sizes are considered to be $G_R \geq 2N_R$ and $G_{T,u} \geq 2N_{T,u}$ \cite{rodriguez2018frequency}. The frequency-independent phase shifters as shown in Eqs. \eqref{multi_freq} are modeled as \cite{venugopal2017channel}
\vspace{-4mm}
\begin{align}
    \mathbf{F}_{\text{RF},m,u}(\kappa,\ell)& = \frac{1}{\sqrt{N_{T,u}}} \text{exp}(j\varnothing_{\kappa,\ell}),\\
    \mathbf{W}_{\text{RF},m}(\kappa,\ell) &= \frac{1}{\sqrt{N_R}}\text{exp}(j\vartheta_{\kappa,\ell}),
\end{align}
where the phases $\varnothing_{\kappa,\ell}, \vartheta_{\kappa,\ell}$ are randomly sampled with uniform probability over $\mathcal{T} = \Big\{0,\frac{2 \pi}{2^{N_Q}}, \cdots, \frac{\left(2^{N_Q} - 1\right)}{2^{N_Q}}\Big\}.$
For performance evaluation, we employ the widely used NMSE metric given
\begin{align}
    \text{NMSE} = \frac{\sum_{k=0}^{K-1} \parallel \widehat{\mathbf{H}}[k] - \mathbf{H}[k] \parallel_F^2}{\sum_{k=0}^{K-1} \parallel \mathbf{H}[k] \parallel_F^2}.
\end{align}
\vspace{-4mm}
\subsection{GMM based MU AoA/ AoD generation}
We harness a GMM for generating the AoAs/AoDs, as it provides an accurate representation of the angular characteristics in the THz band \cite{lin2015adaptive}. To generate angles corresponding to distinct users, the mean angle for each user is drawn from a uniform distribution, $\tilde{\theta}_u \sim \mathcal{U}[-180^\circ, 180^\circ]$, where $\tilde{\theta}_u$ denotes the mean angle of the $u$-th user. Around each mean, the angular spread is modeled by a Gaussian kernel of variance $\check{\sigma}^2$, where $\check{\sigma}^2$ is chosen according to the measurements reported in \cite{priebe2011aoa} (Table III). Thus, the AoA/AoD distribution using a two-component GMM can be expressed as
\begin{align}
    \text{GMM}(x_u) = \sum_{\iota=1}^2 a_{u,\iota} \mathcal{N}(x_u|\tilde{\theta}_{u,\iota}, \check{\sigma}^2),
\end{align}
where $a_{u,\iota}$ represent the mixture weight. They are randomly drawn from a uniform distribution $\{a_{u,1},a_{u,2}\} \sim \mathcal{U}(0,1)$ satisfying $a_{u,1}+a_{u,2} = 1,$ ensuring a valid convex combination. Furthermore, to avoid angular overlap among different users, we impose a \textit{minimum separation constraint of} 
\begin{align}
    |\tilde{\theta}_u - \tilde{\theta}_v| \geq d_{\text{min}} \, \forall \, u \neq v
\end{align}
where $d_{\text{min}}$ is chosen to guarantee sufficiently distinct spatial signatures. For our simulations we have considered $d_{\text{min}} \approx 5^\circ$. This ensures that although all users share the same angular spread parameter $\sigma,$ their mean directions remain sufficiently separated.
\begin{figure*}
\centering
\subfloat[]{\includegraphics[scale=0.31]{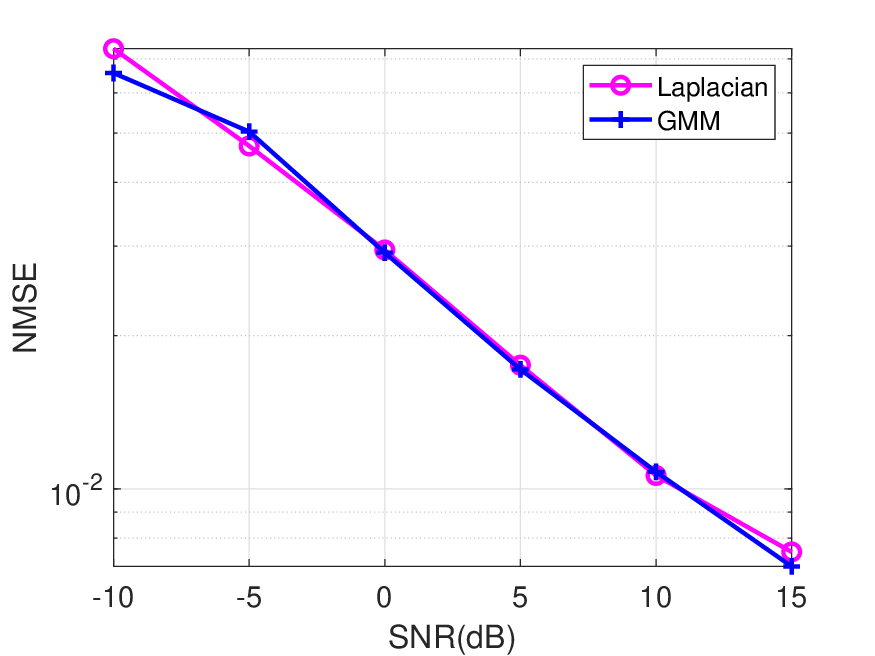}}
	\hfil
	\hspace{-5pt}\subfloat[]{\includegraphics[scale=0.31]{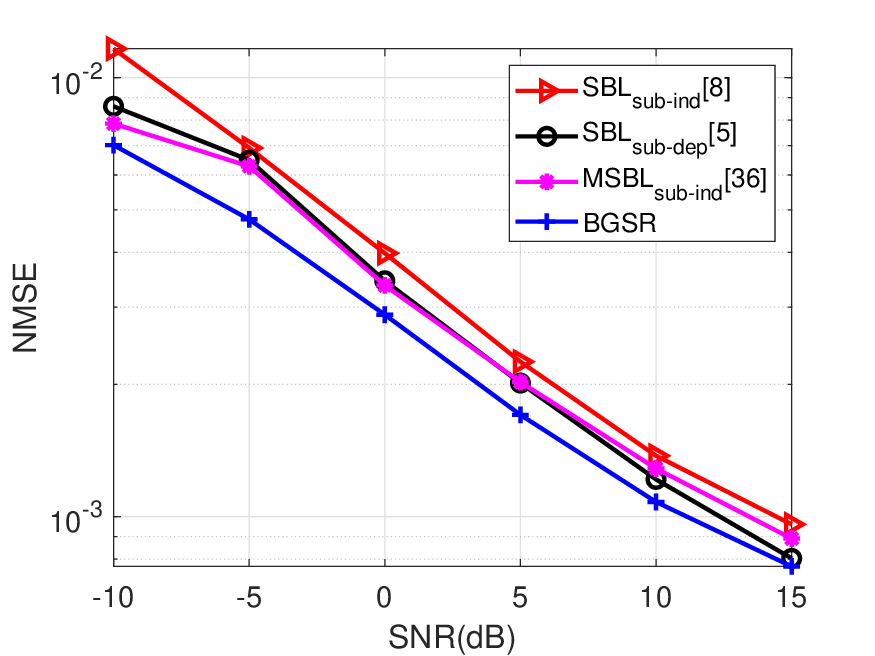}}
    \hfil
\hspace{-5pt}\subfloat[]{\includegraphics[scale=0.31]{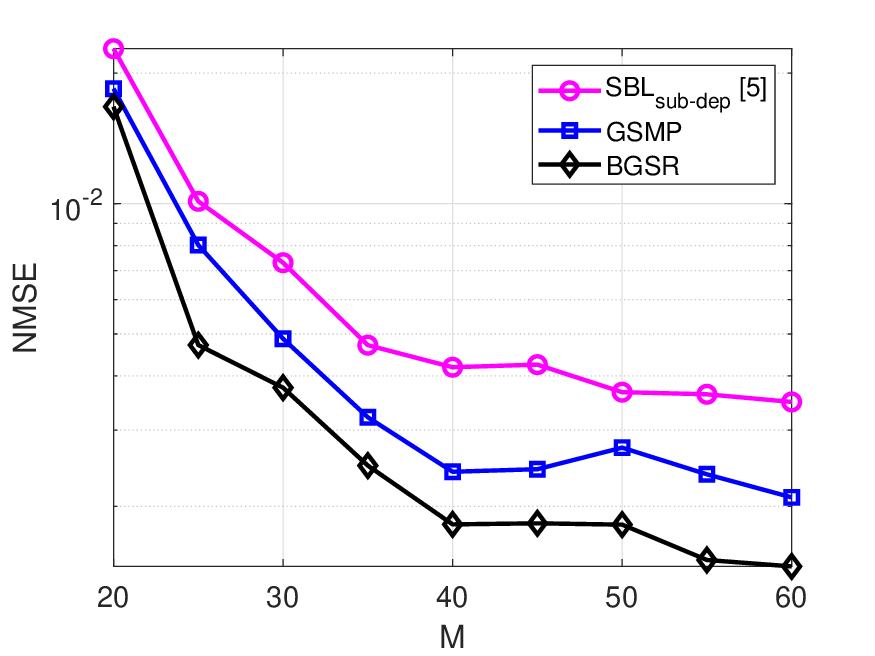}}
	\hfil
	\hspace{-5pt}\subfloat[]{\includegraphics[scale=0.31]{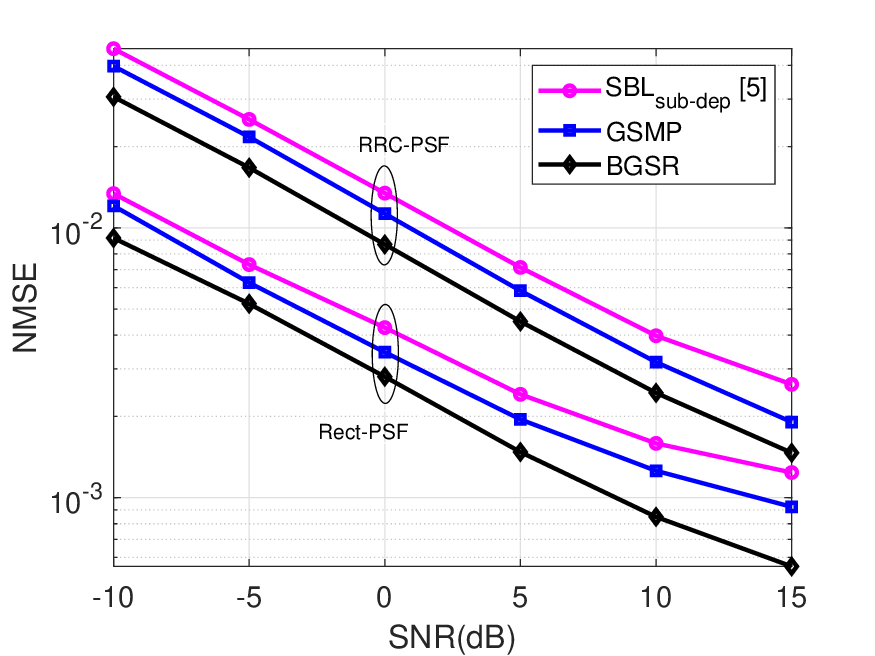}}
\vspace{-2mm}
	\caption{$(a)$ NMSE vs. SNR comparison for the proposed BGSR framework with AoA/AoD generated from Laplacian and Gaussian mixture distributions $(b)$ NMSE vs SNR comparison for subcarrier-independent and subcarrier-dependent Bayesian frameworks $ \left(c\right) $ NMSE vs training frames for MU scenario  $ \left(d\right) $ NMSE vs SNR performance comparison for the GSMP and BGSR based sparse estimation techniques with RRC-PSF and Rect-PSF based dual-wideband channel for MU scenario}
\label{new_plots}
\end{figure*}
\subsection{MU performance analysis}
In this section, we evaluate the performance of the proposed BGSR based CSI estimation technique. The parameters considered for simulation are shown in Table-\ref{simulation_para}. Fig. \ref{MU_BCRLB}(a) portrays the improved performance of the proposed BGSR technique in contrast to several orthodox sparse estimation techniques, such as SBL \cite{tipping2001sparse}, OMP \cite{cheng2016matrix}, MFOCUSS \cite{cotter2005sparse}, GSMP for THz hybrid MIMO systems. The performance of the OMP algorithm is significantly affected by the choice of the stopping criterion and the dictionary matrix, as well as error propagation. Additionally, convergence-related issues are experienced with the MFOCUSS algorithm, which exhibits high sensitivity to variations in the regularization parameter. The SMV-based SBL does not exploit group sparsity, as it estimates the channel separately on individual subcarriers, which degrades its performance. The GSMP framework demonstrates noteworthy performance improvement, since it leverages the group sparsity via the estimation of the channel over all the subcarriers. However, similar to the OMP algorithm, the GSMP framework is sensitive to the stopping criterion and to the choice of the dictionary matrix. By contrast, the proposed BGSR framework leverages EM-based hyperparameter estimation, eliminating the need for manual parameter adjustments, while ensuring robust convergence. Furthermore, the proposed BGSR framework is unaffected by the selection of a dictionary matrix and stopping threshold, which vindicates the efficacy of the proposed algorithm. Moreover, observe from the figure that the performance of the proposed BGSR framework closely tracks the BCRB. This is of significant importance, as the BCRB is derived under the assumption of perfect knowledge of the AoA/ AoDs, which is an ideal scenario, while the proposed BGSR framework does not require any prior information. Note that we have assumed $\mathbf{C}_w$ to be perfectly known. However, as depicted in Fig. \ref{MU_BCRLB}(b), if the noise covariance is estimated, the NMSE performance remains almost identical, illustrating that the proposed BGSR algorithm is robust to noise mismatches.

Fig. \ref{MU_BCRLB}(c) shows the accuracy of the data detection achieved using the MMSE receiver for the BGSR and other traditional sparse estimation based approaches along with that of the Genie-aided detector. It is evident from the figure that the resultant BER improves upon increasing the data SNR and yields the best performance for the proposed BGSR based technique. This is in line with the trend for the NMSE of the estimated CSI. Furthermore, the BERs of the BGSR and GSMP techniques closely approximate the hypothetical Genie-aided detector, demonstrating the efficacy of the proposed schemes.  Fig. \ref{MU_BCRLB}(d) illustrates the impact of a diffused multipath channel on the proposed BGSR scheme for $N_{ray} = \left\{3,4\right\}$, with a comparison to the GSMP approach. As the number of diffused rays increases, the performance of both schemes degrades. This behavior arises from the GMM adopted for generating AoAs/AoDs. In the GMM framework, additional rays correspond to extra Gaussian components, which broaden the angular spread around each user’s mean direction. A larger spread reduces the sparsity of the beamspace channel, as more angular bins become active. Consequently, the dictionary atoms become less distinguishable, which diminishes the efficacy of sparse recovery and leads to degraded NMSE performance. Note that, in practical THz scenarios, the true angular statistics may deviate from the Gaussian assumption. As shown in Fig. \ref{new_plots}(a), the BGSR framework exhibits almost identical NMSE performance under both the Gaussian and non-Gaussian distributions. This robustness arises because the Bayesian framework primarily exploits the inherent sparsity of the beamspace channel, rather than relying on the \textit{exact functional form of the angular distribution}. While the Gaussian prior is adopted in our formulation due to its analytical tractability and closed-form EM updates, the results confirm that the algorithm remains resilient to moderate distributional mismatches. Performance degradation is observed only when the variance of rays becomes significantly larger or when the number of clusters increase significantly, as this reduces sparsity and leads to overlapping dictionary atoms, as depicted in Fig. \ref{MU_BCRLB}(d). These results demonstrate that the BGSR framework is not limited to Gaussian angular models and generalizes well to other practical distributions. 
\begin{figure*}
	\centering
	\hspace{-5pt}\subfloat[]{\includegraphics[scale=0.41]{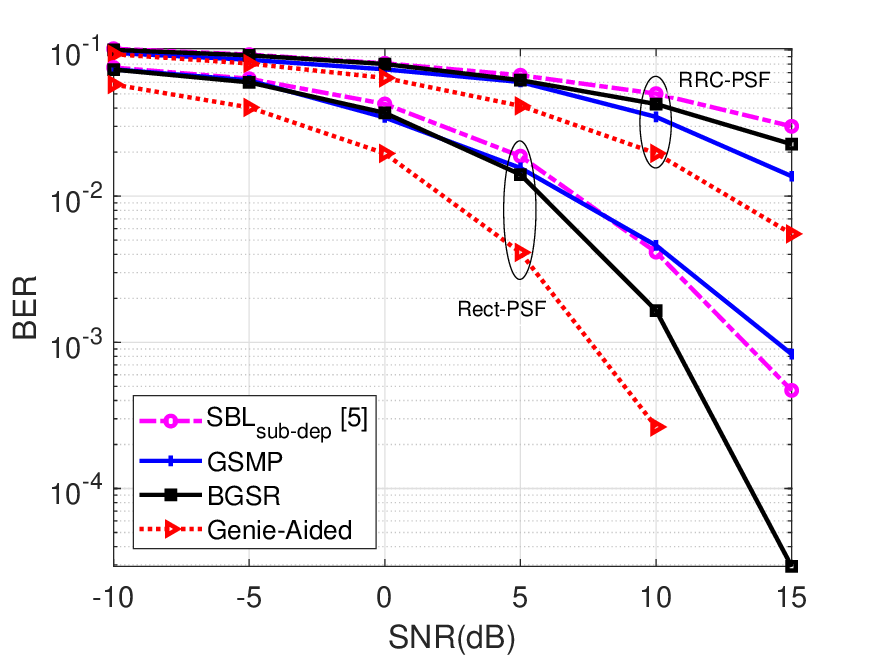}}
        \hfil
	\hspace{-5pt}\subfloat[]{\includegraphics[scale=0.41]{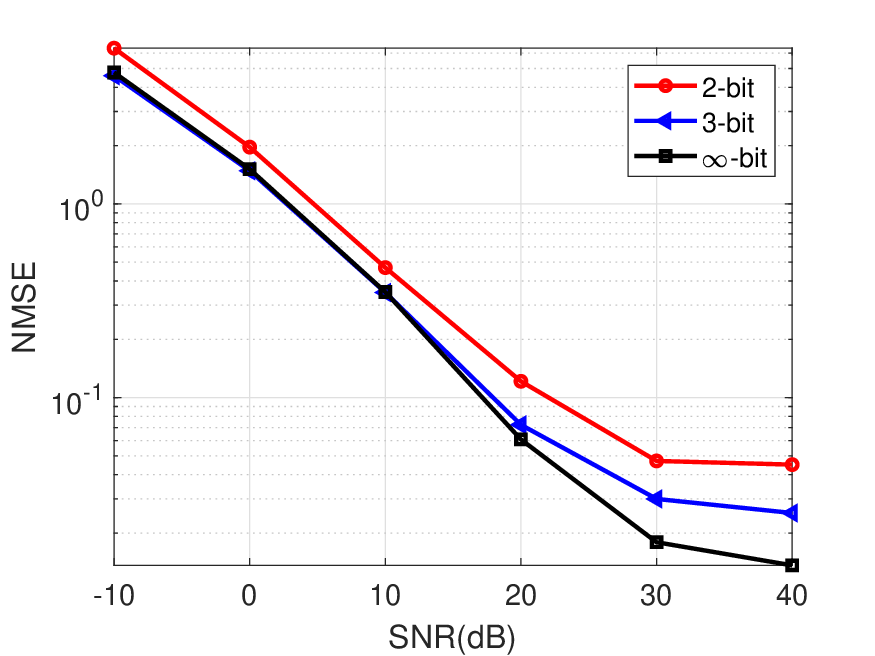}}
    \hfil
	\hspace{-5pt}\subfloat[]{\includegraphics[scale=0.41]{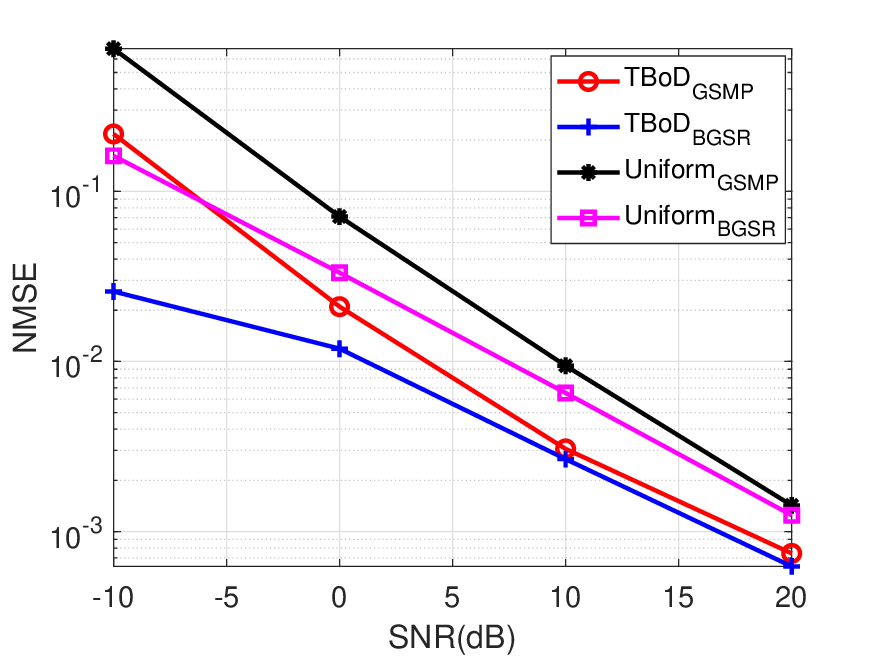}}
 \vspace{-2mm}
	\caption{$ \left(a\right) $ BER vs SNR performance comparison for the GSMP and BGSR based sparse estimation techniques with RRC-PSF and Rect-PSF based dual-wideband channel for MU scenario $ \left(b\right) $ Effect of low resolution ADCs for the BGSR based technique for MU scenario. $\left(c\right)$ NMSE vs SNR performance comparison for uniform and TBoD based channel estimation} \vspace{-1.2 \baselineskip}
    \label{MU_DW_SW} 
\end{figure*}

Fig. \ref{new_plots}(b) compares the NMSE vs SNR performance of subcarrier-independent and subcarrier-dependent dictionaries-based sparse recovery frameworks. The relatively poor performance of the subcarrier-independent dictionary-based SMV-SBL \cite{tipping2001sparse} is primarily due to its reliance on a dictionary constructed at the central subcarrier, which cannot accurately capture the frequency-dependent variations introduced by the beam-squint effect. This mismatch leads to degraded channel estimation performance across a wideband. When a subcarrier-dependent dictionary is incorporated into the SMV-SBL \cite{garg2024angularly}, the dictionary better reflects the frequency-selective characteristics, resulting in performance improvement. However, the single-measurement formulation remains limited, since it does not exploit the joint sparsity across users. By contrast, the MMV-SBL framework substantially enhances the performance by jointly estimating all user channels under a shared sparsity structure. The subcarrier-independent dictionary-based MMV-SBL \cite{wipf2007empirical} provides noticeable improvement compared to the SMV variants, and the proposed BGSR achieves the best performance, as it simultaneously leverages joint sparsity and accurately models the beam-squint effect, leading to the lowest NMSE across all SNR values. Fig. \ref{new_plots}(c) depicts the NMSE versus the number of pilot blocks $M$ for the proposed BGSR and GSMP techniques in comparison with traditional Bayesian methods. The NMSE decreases with increasing number of pilot blocks, as a larger number of pilots provides richer observations and thereby improves channel estimation accuracy. Furthermore, BGSR consistently outperforms the other schemes by leveraging simultaneous sparsity within the Bayesian framework, while GSMP achieves moderate performance and SBL performs the worst due to its limited ability to exploit the joint sparsity structure. These results further validate the robustness of the proposed BGSR and GSMP frameworks.

Fig. \ref{new_plots}(d) compares the performance of the BGSR and GSMP based channel estimation approaches by considering both the RRC-PSF and Rect-PSF based dual-wideband channels. The Rect-PSF based dual-wideband channel suffers from spectral leakage, which leads to a higher level of interference between the adjacent channels, which may erode the overall performance of the communication system. On the other hand, the RRC-PSF based dual wideband channel introduces the problem of signal strength reduction at the edges of the frequency band. This attenuation may adversely affect signal recovery at the receiver end, as the diminished signal may fall below the receiver’s detection threshold. Thus, there is a clear-trade off between both the models and the consideration of the filter depends upon the specific application. Fig. \ref{MU_DW_SW}(a) compares the performance of the RRC-PSF and Rect-PSF based dual-wideband channels in terms of its BER. The improved accuracy of GSMP and BGSR based CSI learning techniques is clearly evident from the figure. Moreover, the BER is compared to that of a Genie-aided hypothetical scenario, which assumes perfect knowledge of the THz hybrid MIMO channel. It is evident that the BGSR technique learns the dual-wideband channel effectively and performs close to the ideal genie receiver. Thus, the improved performance makes the BGSR technique eminently suitable for practical THz system implementation.
\begin{figure*}
\centering
\subfloat[]{\includegraphics[scale=0.31]{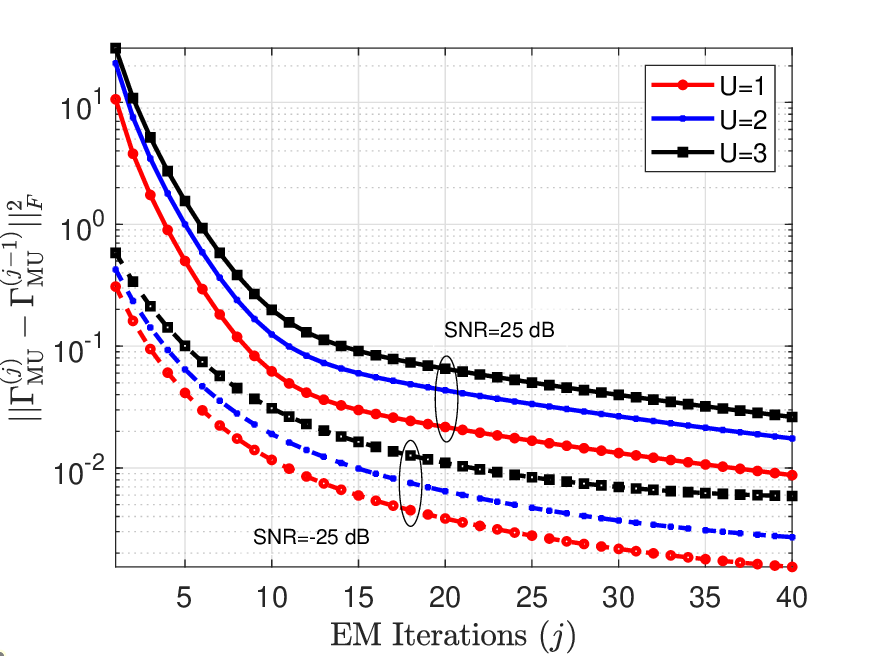}}
	\hfil
	\hspace{-5pt}\subfloat[]{\includegraphics[scale=0.31]{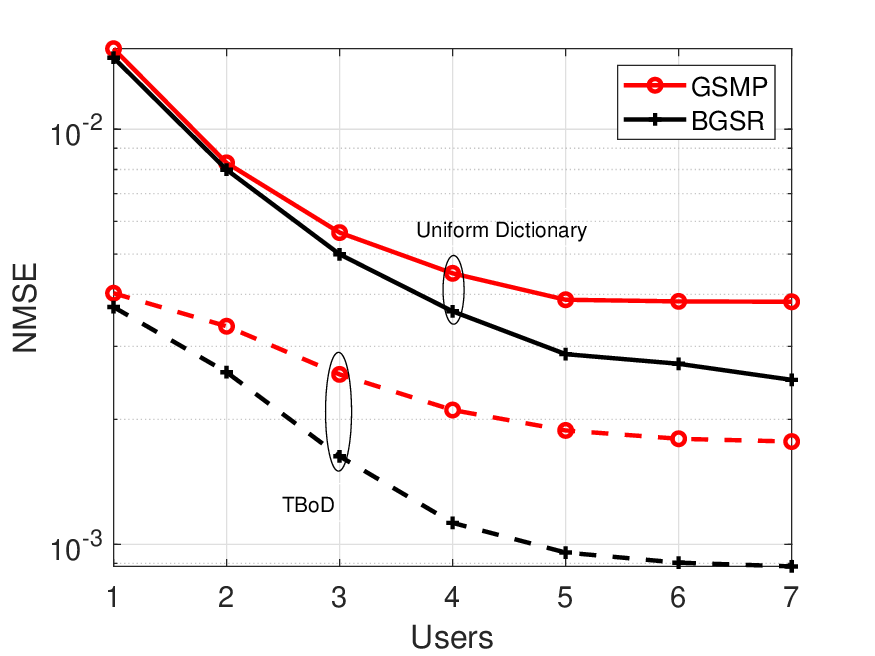}}
    \hfil
\hspace{-5pt}\subfloat[]{\includegraphics[scale=0.31]{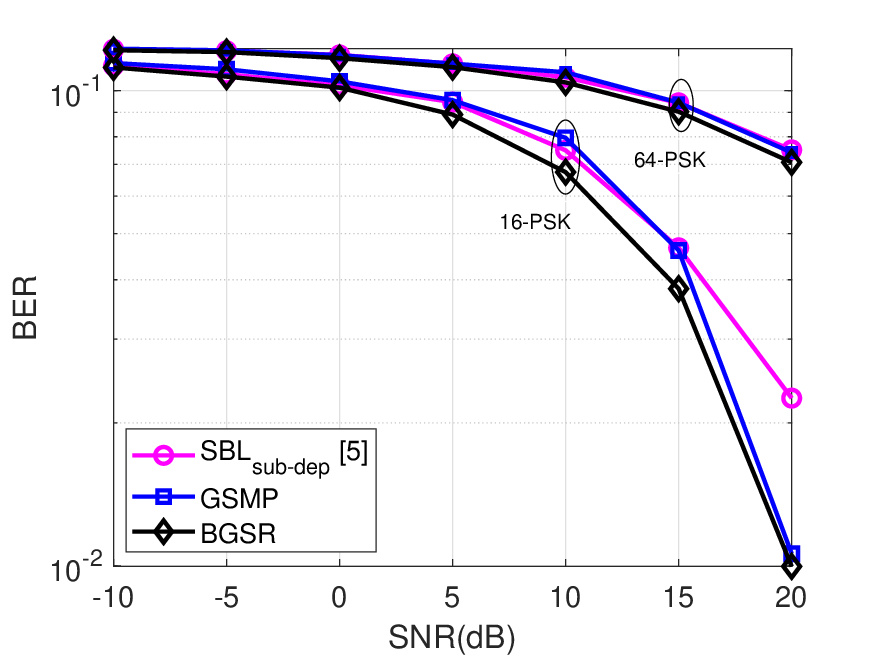}}
	\hfil
	\hspace{-5pt}\subfloat[]{\includegraphics[scale=0.31]{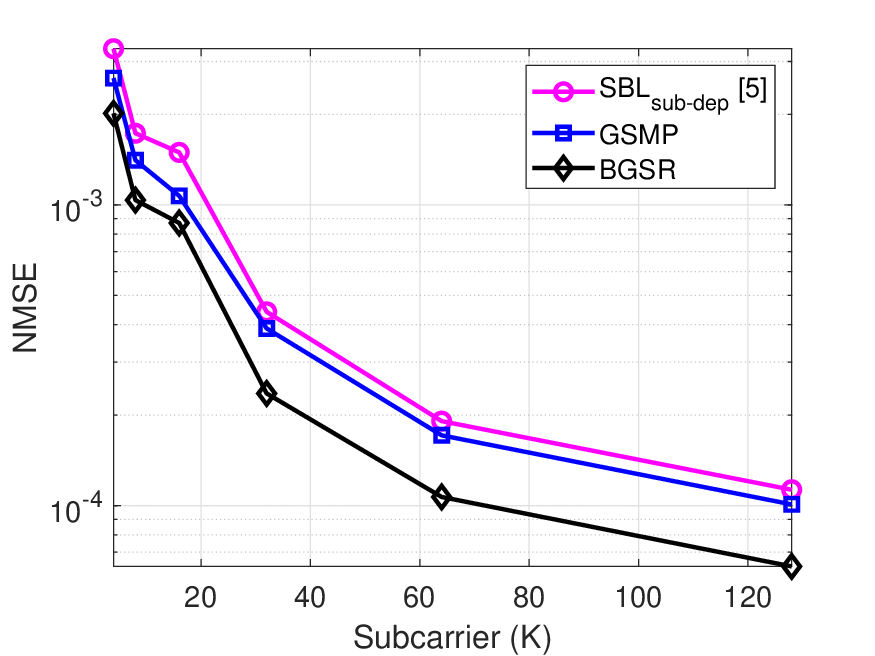}}
\vspace{-2mm}
	\caption{$(a)$ $\parallel \mathbf{\Gamma}_{\text{MU}}^{(j)} - \mathbf{\Gamma}_{\text{MU}}^{(j-1)} \parallel_\mathcal{F}^2$ vs. number of EM iterations by varying the users $(b)$ NMSE vs users for both the proposed algorithms under uniform and TBoD based dictionary constraints. $ \left(c\right) $ BER vs SNR performance comparison for the proposed GSMP and BGSR techniques with different modulation for SU scenario $ \left(d\right) $ NMSE vs subcarrier performance comparison for the GSMP and BGSR based approaches for SU scenario.}
\label{SU_mod} \vspace{-1 \baselineskip}
\end{figure*}

Fig. \ref{MU_DW_SW}(b) demonstrates the influence of using low-resolution ADCs on the estimated CSI performance for the BGSR framework. One can readily observe from the figure that the proposed BGSR technique with $3$-bit ADC resolution closely approaches the performance of an $\infty$-bit resolution transceiver. This observation is of considerable importance as it highlights the practicality of the BGSR scheme, which often necessitates low-resolution ADCs to efficiently handle its substantial bandwidth at a nominal power consumption. Furthermore, none of the existing studies has addressed the influence of low-resolution ADCs in the face of dual-wideband effects. Moreover, it can be observed from Table-\ref{bit-resolution} that the quantization noise-to-signal power ratio $\upsilon$ increases with decreasing the ADC resolution $b$. Therefore, for ADCs with fewer than $3$ bits $(b<3)$, the higher quantization noise leads to significant performance degradation, as illustrated in Fig. \ref{MU_DW_SW}(b). Thus, the proposed model not only conserves power but also leads to a reduction in hardware costs and energy consumption, thereby highlighting the practical advantages of the proposed approach. Fig \ref{MU_DW_SW} (c) compares the performance of the existing on-grid dictionary and of the TBoD matrix. The TBoD outperforms the on-grid dictionary based performance for both the proposed techniques due to its ability to more accurately represent continuous angular domains. Unlike the on-grid approach, which is constrained by discretized grid points and suffers from basis mismatch errors, the off-grid dictionary eliminates such limitations by allowing a finer and more flexible representation of the angular information. This enhanced adaptability leads to improved estimation accuracy and to enhanced overall performance. Fig. \ref{SU_mod}(a) depicts the no. of the convergence of the proposed BGSR technique as a function of EM iterations for multiple users. To achieve a target NMSE of approximately $0.1$, the algorithm requires about $8$ EM iterations for a single user, but roughly $14$ iterations for three users. Although the required no. of EM iterations increases from approximately $8$ to $14$, this growth is not strictly linear with respect to the number of users. The fact that the increment is less than a factor of \textbf{three} indicates that the dictionary-based algorithm benefits from the shared angular structure of multiple users, preventing the computational complexity from growing linearly. Hence, while the MU extension naturally expands the sparse channel space, the iterative process still leverages inter-user correlations, preventing an unmanageable increase in complexity. This sub-linear scaling highlights that the MU scenario remains computationally tractable, enabling the framework to effectively accommodate additional users without requiring an unbounded increase in iteration count. Fig. \ref{SU_mod}(b) compares the NMSE performance for different numbers of users. As the number of users increases, the NMSE performance improve due to the availability of additional measurement vectors, which enhances the estimation accuracy. Moreover, the TBoD-based dictionary outperforms the uniform dictionary, as it provides a more accurate representation of the continuous angular domains. This improvement highlights the advantage of structured dictionary learning in capturing the underlying sparsity of the THz channel, leading to more efficient channel estimation. Therefore, this sub-linear scaling ensures that the framework remains computationally feasible, while efficiently supporting MU in high-dimensional THz systems.
\subsection{Trade-off between communication resources and computational complexity}
As a special case for MU, Fig. \ref{SU_mod}(c) and (d) compares the performance for an SU scenario. All the other parameters are identical to the ones in Table-\ref{simulation_para}. Fig. \ref{SU_mod}(c) compares the BER performance of the proposed channel estimation techniques for $\left\{16, 64\right\}$-PSK modulation. It is apparent from the figure that the proposed technique outperforms the other sparse techniques for higher-order modulation as well. Moreover, the increased constellation size of say $64$-PSK mandates a higher SNR due to its tighter spacing between the constellation points, leading to a higher BER. Fig. \ref{SU_mod}(d) demonstrates our performance comparison between the NMSE and the number of subcarriers. Observe that by increasing the number of subcarriers, a large number of pilot symbols can be embedded within the signal, which further helps in improving the channel estimation performance. Moreover, given the precise channel knowledge at each subcarrier, the receiver can accurately recover the data, thus leading to improved performance. Furthermore, the improved performance of the proposed algorithms is due to the fact that they leverage the unique sparsity of the channel within a group framework, enabling group channel estimation across all the subcarriers. Therefore, as the number of subcarriers $K$ increases, the number of group sparse vectors also increases, enhancing the overall estimation performance. Additionally, the pilots are loaded onto subcarriers, meaning that the number of pilot blocks effectively increases with the number of subcarriers. Furthermore, as the number of pilot blocks increases, the NMSE performance also improves.

A clear trade-off exists between SU and MU channel estimation. SU estimation is computationally efficient due to its lower dimensionality; however, it requires sequential processing of user channels, often over different time slots, and relies on separate pilots for each user. As a result, by the time later users are estimated, the CSI of earlier users may already be outdated, especially in dynamic environments. By contrast, joint estimation is more resource-intensive but leverages a common communication sequence and shared pilot transmission over the same time–frequency band to simultaneously estimate all user channels. This joint processing eliminates sequential delays, ensures consistency across users, and avoids performance loss due to outdated CSI. Moreover, MU estimation leverages the inherent group sparsity and inter-user correlations in THz channels, which cannot be fully exploited by independent SU estimation. Consequently, despite the higher computational cost, joint estimation achieves more reliable, up-to-date, and spectrally efficient CSI acquisition, making it highly advantageous in practical MU systems.
\section{Conclusions} \label{conclusion}
New channel learning techniques have been conceived by exploiting the angular domain sparsity in practical THz hybrid MIMO systems. The THz channel was rigorously modeled considering the absorption coefficient (obtained through the HITRAN database), reflection losses, diffused ray modeling including the dual wideband effect. Furthermore, we also considered the effect of low resolution ADCs along with the practical implementation of THz hybrid MIMO systems. The i/o model was linearized using the popular Bussgang decomposition. Moreover, an SC-FDE based system model was conceived to convert the TD processing into its equivalent FD counterpart, and to provide frame-wise processing for attaining significant performance improvements. Furthermore, a novel Bayesian learning based channel learning framework i.e., BGSR, was proposed whose performance was also characterized by deriving the MU-BCRB. Moreover, the performance of the RRC-PSF and Rect-PSF based dual wideband channels was explicitly compared. Our simulation results offer empirical evidence of the enhanced performance achieved by the BGSR scheme in comparison to other sparse estimation techniques.
\appendices
\section{Calculation of Noise Covariance} \label{noise_cova}
The received signal after performing analog combining and before passing through the low-resolution ADC can be expressed as
\begin{align}
    \mathbf{y}_{m,\text{MU}}(q) = \mathbf{W}_{\text{RF},m}^H \sum_{u=1}^U\mathbf{H}_{q,u} \otimes (\mathbf{F}_{\text{RF},m,u}\mathfrak{a}_{m,u}^{(q)})+\mathbf{W}_{\text{RF},m}^H\tilde{\mathbf{v}}_m(q), \label{before-ADC}
\end{align}
Let $\mathbf{x}_{m,u}(q) = \mathbf{F}_{\text{RF},m,u}\mathfrak{a}_{m,u}^{(q)}$ for notational simplicity. Moreover, $\mathbb{E}(\mathbf{x}_{m,u}(q)\mathbf{x}_{m,u}^H(q)) = \sigma_b^2 \mathbf{F}_{\text{RF},m,u} \mathbf{F}_{\text{RF},m,u}^H = \mathbf{R}_{xx,m}$. Expanding the circular convolution as defined in Section-\ref{notan}, Eq. \eqref{before-ADC} can be re-written as
\begin{align}
    &\tilde{\mathbf{y}}_m(q) = \mathbf{W}_{\text{RF},m}^H\sum_{u=1}^U[\mathbf{H}_u(0)\mathbf{x}_{m,u}(q)+ \mathbf{H}_u(1)\mathbf{x}_{m,u}(q-1)+ \notag \\ & \cdots+ \mathbf{H}_u(K-1)\mathbf{x}_{m,u}(q-K+1)] + \mathbf{W}_{\text{RF},m}^H\tilde{\mathbf{v}}_m(q). \label{expand_ADC}
\end{align}
Let $\tilde{\mathbf{C}}_m \in \mathbb{C}^{N_{\text{RF}}^R \times N_{\text{RF}}^R}$ represents the covariance matrix which is defined as $\tilde{\mathbf{C}}_m = \mathbb{E}\{\tilde{\mathbf{y}}_m(q)\tilde{\mathbf{y}}_m^H(q)\}$. After substituting $\tilde{\mathbf{y}}_m(q)$ into the expression of the covariance matrix, we obtain
\begin{align}
    \tilde{\mathbf{C}}_m & = \mathbf{W}_{\text{RF},m}^H \sum_{u=1}^U[\mathbf{H}_u(0)\mathbf{R}_{xx,m}\mathbf{H}_u^H(0)+\cdots+  \notag \\ & \mathbf{H}_u(K-1)\mathbf{R}_{xx,m}\mathbf{H}_u^H(K-1)]\mathbf{W}_{\text{RF},m}^H + \sigma_n^2\mathbf{W}_{\text{RF},m}^H\mathbf{W}_{\text{RF},m}.
\end{align}
Therefore, the received signal covariance matrix before passing through low-resolution ADCs can be expressed as $\tilde{\mathbf{C}}_m = \mathbf{W}_{\text{RF},m}^H\mathbf{Q}_m\mathbf{W}_{\text{RF},m} + \sigma_n^2\mathbf{W}_{\text{RF},m}^H\mathbf{W}_{\text{RF},m}$ where $\mathbf{Q}_m = \sum_{u=1}^U \sum_{n=0}^{K-1}\mathbf{H}_u(n)\mathbf{R}_{xx,m}\mathbf{H}_u^H(n)$. Additionally, the noise covariance matrix after passing through low-resolution ADC is given as $\mathbf{C}_m = \varepsilon(1-\varepsilon)\mathrm{diag}(\tilde{\mathbf{C}}_m)$.
\bibliographystyle{IEEEtran}
\bibliography{References}

@article{wang2024knowledge,
  title={{Knowledge and Data Dual-Driven Channel Estimation and Feedback for Ultra-Massive MIMO Systems under Hybrid Field Beam Squint Effect}},
  author={Wang, Kuiyu and Gao, Zhen and Chen, Sheng and Ning, Boyu and Chen, Gaojie and Su, Yu and Wang, Zhaocheng and Poor, H Vincent},
  journal={IEEE Transactions on Wireless Communications},
  year={2024},
  publisher={IEEE}
}

@article{fan2015uplink,
  title={{Uplink achievable rate for massive MIMO systems with low-resolution ADC}},
  author={Fan, Li and Jin, Shi and Wen, Chao-Kai and Zhang, Haixia},
  journal={IEEE Communications Letters},
  volume={19},
  number={12},
  pages={2186--2189},
  year={2015},
  publisher={IEEE}
}

@article{wipf2007empirical,
  title={An empirical Bayesian strategy for solving the simultaneous sparse approximation problem},
  author={Wipf, David P and Rao, Bhaskar D},
  journal={IEEE Transactions on Signal Processing},
  volume={55},
  number={7},
  pages={3704--3716},
  year={2007},
  publisher={IEEE}
}

@article{li2024hybrid,
  title={Hybrid near-and far-field THz UM-MIMO channel estimation: a sparsifying matrix learning-aided Bayesian approach},
  author={Li, Yuanjian and Madhukumar, AS},
  journal={IEEE Transactions on Wireless Communications},
  year={2024},
  publisher={IEEE}
}

@article{sarieddeen2021overview,
  title={{An overview of signal processing techniques for terahertz communications}},
  author={Sarieddeen, Hadi and Alouini, Mohamed-Slim and Al-Naffouri, Tareq Y},
  journal={Proceedings of the IEEE},
  volume={109},
  number={10},
  pages={1628--1665},
  year={2021},
  publisher={IEEE}
}

@article{mo2017channel,
  title={{Channel estimation in broadband millimeter wave MIMO systems with few-bit ADCs}},
  author={Mo, Jianhua and Schniter, Philip and Heath, Robert W},
  journal={IEEE Transactions on Signal Processing},
  volume={66},
  number={5},
  pages={1141--1154},
  year={2017},
  publisher={IEEE}
}

@article{wang2018spatial,
  title={Spatial-and frequency-wideband effects in millimeter-wave massive {MIMO} systems},
  author={Wang, Bolei and Gao, Feifei and Jin, Shi and Lin, Hai and Li, Geoffrey Ye},
  journal={IEEE Transactions on Signal Processing},
  volume={66},
  number={13},
  pages={3393--3406},
  year={2018},
  publisher={IEEE}
}

@article{tipping2001sparse,
  title={{Sparse Bayesian learning and the relevance vector machine}},
  author={Tipping, Michael E},
  journal={Journal of Machine Learning Research},
  pages={211--244},
  year={2001}
}

@article{wipf2004sparse,
  title={Sparse {B}ayesian learning for basis selection},
  author={Wipf, David P and Rao, Bhaskar D},
  journal={IEEE Transactions on Signal Processing},
  volume={52},
  pages={2153--2164},
  year={2004},
  publisher={IEEE}
}

@article{lin2015adaptive,
  title={Adaptive beamforming with resource allocation for distance-aware multi-user indoor {TeraHertz} communications},
  author={Lin, Cen and Li, Geoffrey Ye},
  journal={IEEE Transactions on Communications},
  volume={63},
  number={8},
  pages={2985--2995},
  year={2015},
  publisher={IEEE}
}

@article{li2020dynamic,
  title={Dynamic hybrid beamforming with low-resolution {PSs} for wideband {mmWave MIMO-OFDM} systems},
  author={Li, Hongyu and Li, Ming and Liu, Qian and Swindlehurst, A Lee},
  journal={IEEE Journal on Selected Areas in Communications},
  volume={38},
  number={9},
  pages={2168--2181},
  year={2020},
  publisher={IEEE}
}

@article{ding2018bayesian,
  title={Bayesian channel estimation algorithms for massive {MIMO} systems with hybrid analog-digital processing and low-resolution {ADCs}},
  author={Ding, Yacong and Chiu, Sung-En and Rao, Bhaskar D},
  journal={IEEE Journal of Selected Topics in Signal Processing},
  volume={12},
  number={3},
  pages={499--513},
  year={2018},
  publisher={IEEE}
}

@article{cheng2016matrix,
  title={Matrix-inversion-free compressed sensing with variable orthogonal multi-matching pursuit based on prior information for {ECG} signals},
  author={Cheng, Yih-Chun and Tsai, Pei-Yun and Huang, Ming-Hao},
  journal={IEEE Transactions on Biomedical Circuits and Systems},
  volume={10},
  number={4},
  pages={864--873},
  year={2016},
  publisher={IEEE}
}

@article{huang2018channel,
  title={Channel estimation in {MIMO-OFDM} systems based on a new adaptive greedy algorithm},
  author={Huang, Yuan and He, Yigang and Luo, Qiwu and Shi, Luqiang and Wu, Yuting},
  journal={IEEE Wireless Communications Letters},
  volume={8},
  number={1},
  pages={29--32},
  year={2018},
  publisher={IEEE}
}

@inproceedings{du2015mixed,
  title={Mixed norm minimization for {MIMO} cellular interference channel},
  author={Du, Huiqin and Ratnarajah, Tharm and Sellathurai, Mathini and Chambers, Jonathon},
  booktitle={2015 IEEE 16th International Workshop on Signal Processing Advances in Wireless Communications (SPAWC)},
  pages={635--639},
  year={2015},
  organization={IEEE}
}

@inproceedings{chethan2016iterative,
  title={An iterative re-weighted minimization framework for resource allocation in the single-cell relay-enhanced {OFDMA} network},
  author={Chethan, Kumar A and Murthy, Chandra R},
  booktitle={2016 IEEE 17th International Workshop on Signal Processing Advances in Wireless Communications (SPAWC)},
  year={2016}
}

@article{cotter2005sparse,
  title={Sparse solutions to linear inverse problems with measurement vectors},
  author={Cotter, Shane F and Rao, Bhaskar D and Engan, Kjersti and Kreutz-Delgado, Kenneth},
  journal={IEEE Transactions on Signal Processing},
  volume={53},
  number={7},
  pages={2477--2488},
  year={2005},
  publisher={IEEE}
}

@inproceedings{joseph2015online,
  title={Online recovery of temporally correlated sparse signals using multiple measurement vectors},
  author={Joseph, Geethu and Murthy, Chandra R and Prasad, Ranjitha and Rao, Bhaskar D},
  booktitle={2015 IEEE GLOBECOM},
  pages={1--6},
  year={2015},
  organization={IEEE}
}

@article{sha2021channel,
  title={Channel estimation and equalization for {TeraHertz} receiver with {RF} impairments},
  author={Sha, Ziyuan and Wang, Zhaocheng},
  journal={IEEE Journal on Selected Areas in Communications},
  volume={39},
  number={6},
  pages={1621--1635},
  year={2021},
  publisher={IEEE}
}

@article{srivastava2022hybrid,
  title={Hybrid transceiver design for {TeraHertz MIMO} systems relying on {Bayesian} learning aided sparse channel estimation},
  author={Srivastava, Suraj and Tripathi, Ajeet and Varshney, Neeraj and Jagannatham, Aditya K and Hanzo, Lajos},
  journal={IEEE Transactions on Wireless Communications},
  year={2022},
  publisher={IEEE}
}

@article{srivastava2021data,
  title={Data aided quasistatic and doubly-selective {CSI} estimation using affine-precoded superimposed pilots in millimeter wave {MIMO-OFDM} systems},
  author={Srivastava, Suraj and Nath, Jaitra and Jagannatham, Aditya K},
  journal={IEEE Transactions on Vehicular Technology},
  volume={70},
  number={7},
  pages={6983--6998},
  year={2021},
  publisher={IEEE}
}

@article{venugopal2017channel,
  title={Channel estimation for hybrid architecture-based wideband millimeter wave systems},
  author={Venugopal, Kiran and Alkhateeb, Ahmed and Prelcic, Nuria Gonz{\'a}lez and Heath, Robert W},
  journal={IEEE Journal on Selected Areas in Communications},
  volume={35},
  number={9},
  pages={1996--2009},
  year={2017},
  publisher={IEEE}
}

@article{dovelos2021channel,
  title={Channel estimation and hybrid combining for wideband {TeraHertz} massive {MIMO} systems},
  author={Dovelos, Konstantinos and Matthaiou, Michail and Ngo, Hien Quoc and Bellalta, Boris},
  journal={IEEE Journal on Selected Areas in Communications},
  volume={39},
  number={6},
  pages={1604--1620},
  year={2021},
  publisher={IEEE}
}

@article{jornet2011channel,
  title={Channel modeling and capacity analysis for electromagnetic wireless nanonetworks in the {TeraHertz} band},
  author={Jornet, Josep Miquel and Akyildiz, Ian F},
  journal={IEEE Transactions on Wireless Communications},
  pages={3211--3221},
  year={2011},
  publisher={IEEE}
}

@article{chou2023compressed,
  title={Compressed {Training} for {Dual-Wideband Time-Varying Sub-Terahertz Massive MIMO}},
  author={Chou, Tzu-Hsuan and Michelusi, Nicol{\`o} and Love, David J and Krogmeier, James V},
  journal={IEEE Transactions on Communications},
  year={2023},
  publisher={IEEE}
}

@article{piesiewicz2007scattering,
  title={Scattering analysis for the modeling of {THz} communication systems},
  author={Piesiewicz, Radoslaw and Jansen, Christian and Mittleman, Daniel and Kleine-Ostmann, Thomas and Koch, Martin and Kurner, Thomas},
  journal={IEEE Transactions on Antennas and Propagation},
  volume={55},
  number={11},
  pages={3002--3009},
  year={2007},
  publisher={IEEE}
}

@article{rodriguez2018frequency,
  title={Frequency-domain compressive channel estimation for frequency-selective hybrid millimeter wave {MIMO} systems},
  author={Rodr{\'\i}guez-Fern{\'a}ndez, Javier and Gonz{\'a}lez-Prelcic, Nuria and Venugopal, Kiran and Heath, Robert W},
  journal={IEEE Transactions on Wireless Communications},
  pages={2946--2960},
  year={2018},
  publisher={IEEE}
}

@article{gonzalez2018channel,
  title={Channel estimation and hybrid precoding for frequency selective multiuser {mmWave MIMO} systems},
  author={Gonz{\'a}lez-Coma, Jos{\'e} P and Rodriguez-Fernandez, Javier and Gonzalez-Prelcic, Nuria and Castedo, Luis and Heath, Robert W},
  journal={IEEE Journal of Selected Topics in Signal Processing},
  pages={353--367},
  year={2018},
  publisher={IEEE}
}

@inproceedings{srivastava2020msbl,
  title={{MSBL}-based simultaneous sparse channel estimation in {SC} wideband {mmWave} hybrid {MIMO} systems},
  author={Srivastava, Suraj and Jagannatham, Aditya K},
  booktitle={IEEE Global Communications Conference},
  pages={1--6},
  year={2020},
  organization={IEEE}
}

@article{prasad2015joint,
  title={Joint channel estimation and data detection in {MIMO-OFDM} systems: {A} sparse {Bayesian} learning approach},
  author={Prasad, Ranjitha and Murthy, Chandra R and Rao, Bhaskar D},
  journal={IEEE Transactions on Signal Processing},
  year={2015},
  publisher={IEEE}
}

@article{rothman2009hitran,
  title={The {HITRAN} 2008 molecular spectroscopic database},
  author={Rothman, Laurence S and Gordon, Iouli E and Barbe, Alain and Benner, D Chris and Bernath, Peter F and Birk, Manfred and Boudon, Vincent and Brown, Linda R and Campargue, Alain and Champion, J-P and others},
  journal={Journal of Quantitative Spectroscopy and Radiative Transfer},
  pages={533--572},
  year={2009},
  publisher={Elsevier}
}

@article{ekanadham2011recovery,
  title={{Recovery of sparse translation-invariant signals with continuous basis pursuit}},
  author={Ekanadham, Chaitanya and Tranchina, Daniel and Simoncelli, Eero P},
  journal={IEEE transactions on signal processing},
  volume={59},
  number={10},
  pages={4735--4744},
  year={2011},
  publisher={IEEE}
}

@inproceedings{mezghani2012capacity,
  title={Capacity lower bound of {MIMO} channels with output quantization and correlated noise},
  author={Mezghani, Amine and Nossek, Josef A},
  booktitle={Proc. IEEE Int. Symp. Inf. Theory},
  pages={1--5},
  year={2012}
}

@inproceedings{liu2018hybrid,
  title={Hybrid beamforming for {mmWave} {MIMO-OFDM} system with beam squint},
  author={Liu, Bin and Tan, Weiqiang and Hu, Han and Zhu, Hongbo},
  booktitle={2018 IEEE 29th Annual International Symposium on Personal, Indoor and Mobile Radio Communications (PIMRC)},
  pages={1422--1426},
  year={2018},
  organization={IEEE}
}

@article{piesiewicz2007properties,
  title={Properties of building and plastic materials in the {THz} range},
  author={Piesiewicz, Radoslaw and Jansen, Christian and Wietzke, S and Mittleman, Daniel and Koch, Martin and K{\"u}rner, Thomas},
  journal={International Journal of Infrared and Millimeter Waves},
  year={2007},
  publisher={Springer}
}

@inproceedings{priebe2011aoa,
  title={{AoA, AoD and ToA characteristics of scattered multipath clusters for THz indoor channel modeling}},
  author={Priebe, Sebastian and Jacob, Martin and Kuerner, Thomas},
  booktitle={17th European Wireless -Sustainable Wireless Technologies},
  pages={1--9},
  year={2011},
}

@article{kim2021spatial,
  title={{Spatial wideband channel estimation for mmWave massive MIMO systems with hybrid architectures and low-resolution ADCs}},
  author={Kim, In-Soo and Choi, Junil},
  journal={IEEE Transactions on Wireless Communications},
  volume={20},
  number={6},
  pages={4016--4029},
  year={2021},
  publisher={IEEE}
}

@article{zhang2021analysis,
  title={Analysis of indoor {THz} communication systems with finite-bit {DACs} and {ADCs}},
  author={Zhang, Yujiao and Li, Dan and Qiao, Deli and Zhang, Lei},
  journal={IEEE Transactions on Vehicular Technology},
  volume={71},
  number={1},
  pages={375--390},
  year={2021},
  publisher={IEEE}
}

@inproceedings{nikbakht2021terahertz,
  title={{Terahertz transmit beamforming with 1-bit DACs and ADCs}},
  author={Nikbakht, Rasoul and Lozano, Angel},
  booktitle={2021 29th European Signal Processing Conference (EUSIPCO)},
  pages={826--830},
  year={2021},
  organization={IEEE}
}

@inproceedings{mollen2016one,
  title={{One-bit ADCs in wideband massive MIMO systems with OFDM transmission}},
  author={Mollen, Christopher and Choi, Junil and Larsson, Erik G and Heath, Robert W},
  booktitle={2016 IEEE International Conference on Acoustics, Speech and Signal Processing (ICASSP)},
  pages={3386--3390},
  year={2016},
  organization={IEEE}
}

@article{garg2024angularly,
  title={{Angularly Sparse Channel Estimation in Dual-Wideband Tera-Hertz (THz) Hybrid MIMO Systems Relying on Bayesian Learning}},
  author={Garg, Abhisha and Srivastava, Suraj and Yadav, Nimish and Jagannatham, Aditya K and Hanzo, Lajos},
  journal={IEEE Transactions on Communications},
  year={2024},
  publisher={IEEE}
}
\end{document}